\documentclass[lettersize,journal]{IEEEtran}
\IEEEoverridecommandlockouts
\usepackage{caption}
\usepackage{subfigure}
\usepackage{graphics} 
\usepackage{epsfig} 
\usepackage{amsfonts,amssymb}
\usepackage{multirow}
\usepackage{amsmath}
\graphicspath{{figure/}}
\usepackage{subfigure}
\usepackage{graphics}
\usepackage{multirow}
\usepackage{xcolor}
\usepackage{color}
\DeclareMathOperator*{\argmin}{argmin}

\begin{document}

\title{\LARGE \bf Deep Learning for Efficient CSI Feedback in Massive MIMO: Adapting to New Environments and Small Datasets\\

\thanks{Z. Liu, L. Wang and L. Xu are with the School of Computer Science (National Pilot Software Engineering School), Beijing University of Posts and Telecommunications, China (e-mail: {lzyu, liwang, xulianming}@bupt.edu.cn).}
\thanks{Z. Ding is with the Department of Electrical and Computer Engineering, University of California at Davis, USA (e-mail: zding@ucdavis.edu).} 
}
%
%
%

\author{\IEEEauthorblockN{Zhenyu Liu, \emph{Member, IEEE}, Li Wang, \emph{Senior Member, IEEE}, Lianming Xu, and Zhi Ding, \emph{Fellow, IEEE}} 
}

\maketitle

\begin{abstract}

Deep learning (DL)-based channel state information (CSI) feedback has shown promising potential to improve spectrum efficiency in massive MIMO systems. 
However, practical DL approaches require a sizeable CSI dataset for each scenario, and require large storage or updating bandwidth for multiple learned models. To overcome this costly barrier, we develop a solution for efficient training and deployment enhancement of DL-based CSI feedback by exploiting a lightweight translation model to cope with new CSI environments and by proposing novel dataset augmentation based on domain knowledge. Specifically, we first develop a deep unfolding CSI feedback network, SPTM2-ISTANet+, which employs spherical normalization to address the challenge of path loss variation. We also introduce an integration of a trainable measurement matrix and residual CSI recovery blocks within SPTM2-ISTANet+ to improve efficiency and accuracy. Using SPTM2-ISTANet+ as the anchor feedback model, we propose an efficient scenario-adaptive CSI feedback architecture. This new CSI-TransNet exploits a plug-in module for CSI translation consisting of a sparsity aligning function and lightweight DL module to reuse pretrained models in unseen environments. To work with small datasets, we propose a lightweight and general augmentation strategy based on domain knowledge. Test results demonstrate the efficacy and efficiency of the proposed solution for accurate CSI feedback given limited measurements for unseen CSI environments.

\end{abstract}

\begin{IEEEkeywords}
Massive MIMO, CSI feedback, CSI augmentation, deep learning, multi-scenario
\end{IEEEkeywords}

%
\IEEEpeerreviewmaketitle

\section{Introduction}
Modern wireless communication systems have made tremendous strides in utilizing the spatial diversity afforded by multiple-input multiple-output (MIMO) transceivers to improve radio link performance. In particular, massive MIMO systems have shown great promise for delivering high spectrum and energy efficiency for 5G 
wireless systems and beyond. The efficiency of massive MIMO downlink depends on accurate downlink CSI estimates at gNodeB (gNB) for transmission precoding. 
For massive MIMO systems, such feedback data can be substantial because of the large number of antennas and wide bandwidth. This challenge strongly motivates many
research efforts aimed at accurate downlink CSI feedback in frequency division duplex (FDD) systems.

Compressive sensing (CS) approaches can exploit channel properties including low rank or sparsity in spatial \cite{cs5_3} or temporal domain \cite{cs3_3} to reduce the bandwidth
for CSI feedback.  However, CS approaches rely
on the strong sparsity condition which may not strictly hold and can limit their efficacy \cite{ref:csinet}. A number of recent works have explored
deep neural networks (DNNs) for CSI feedback. 
\textcolor{black}{Deep learning (DL) based CSI feedback has shown
good recovery accuracy and time efficiency by
exploiting characteristics such as spatial and spectral coherence \cite{ref:csinet,dl_multires, guo2019convolutional,tang2021dilated, ref:SRNet, wang2020,cui2022,mourya2023}, bi-directional reciprocity \cite{ref:dualphase,ref:canet}, and temporal correlation \cite{ref:csinet-lstm,liu2022,ref:mason}.
However, a significant challenge with DL models is their requirement for extensive channel measurements for training.} Collecting these CSI measurements can be labor-intensive and
quite costly. Furthermore, a single CSI model tends to exhibit poor performance 
when applied to other radio-frequency (RF) environments due to model mismatch. 


\textcolor{black}{The challenge of limited real-time channel measurements necessitates effective data augmentation strategies to reduce labor-intensive measurement costs and prevent model overfitting. Traditional augmentation techniques in image processing, such as flipping, cropping, or rotation \cite{ref:augmentationref}, however, don't align with the physical radio channels.  \cite{guo2022} multiplies the entire CSI matrix by a random value, neglecting the phase shift differences in various multipaths, which degrades the augmentation performance. To our understanding, dedicated CSI data augmentation remains largely uncharted and requires 
domain knowledge of radio propagation physics. } 
Another direction applies generative adversarial network (GAN) as a blackbox
for CSI augmentation \cite{ref:GAN_yang,ref:GAN_ye,ref:GAN_xiao}. 
Ironically, blackbox GAN itself often requires training with a
large number of CSI measurements. Moreover, 
GAN designed for massive MIMO \cite{ref:GAN_xiao} 
can be computationally costly and 
may require billions of floating point operations (FLOPs).

In addition to data augmentation, to improve CSI recovery accuracy 
from limited available data samples,  DNN
should be capable of handling complex CSI features and variations. Generally, 
existing DL-based CSI feedback works \cite{guo2019convolutional,tang2021dilated,cui2022,mourya2023} have demonstrated substantially better performance against indoor channels 
but tend to be less effective for the more complex outdoor channels. 
Furthermore, CSI from wide bandwidth and massive antennas tend to exhibit a high
degree of variation. Path loss alone 
can vary the CSI magnitude by several orders. Such inherent 
CSI characteristics can be particularly troublesome for DL-based 
CSI compression and feedback system trained with 
only a small dataset of measured CSIs that lack broad representation.  
Thus,
the ability to process massive MIMO CSI from a wide variety 
of deployment scenarios is a difficult challenge in
the practical deployment of DL-based CSI feedback schemes.

\textcolor{black}{The principle of transfer learning
leverages cross-task correlations in different scenarios to reduce
training cost.  However, traditional transfer learning involves comprehensive network fine-tuning  \cite{ref:csi_trans2,ref:csi_trans1}. When a User Equipment (UE) transitions to a new environment, the encoder network necessitates an update. Given that CSI feedback encoder networks \cite{ref:csinet,dl_multires, guo2019convolutional,tang2021dilated, ref:SRNet, wang2020,cui2022,mourya2023} can have up to millions of parameters, this implies a significant transmission bandwidth overhead. }
A UE-friendly
design of multi-task, learning-based CSI feedback
has been proposed for the multiple CSI scenarios  \cite{ref:multitask_csi}. 
By exploiting a shared encoder and multiple task-specific decoders, 
the joint feedback architecture of \cite{ref:multitask_csi} can lead to 
a considerable reduction of UE storage use. However, 
a shared encoder can degrade CSI recovery accuracy due to the lack of environment-specific features. Moreover, information in the
pretrained model has not been fully exploited to facilitate 
practical implementation.


\textcolor{black}{To design a scenario-adaptive and high-accuracy 
CSI feedback solution with limited CSI samples and deployment cost,
we develop an efficient DL feedback architecture, which promises enhanced training and communication-efficient deployment even with minimal measured data. We strive to elevate the recovery accuracy and robustness of the CSI feedback network across diverse environments by intertwining an efficient UE-friendly design and a simplified model-driven data augmentation approach rooted in physical insights and domain expertise. Our contributions are 
summarized as follows:
\begin{itemize}
\item Practical deployment of existing DL-based CSI feedback mechanism
can be hindered by insufficient channel measurements needed to 
train large-scale DNNs. Training multiple models against various radio
environments further elevates the difficulty. 
In this paper, we design a scenario-adaptive CSI compression 
framework 
which requires few channel measurements for training and can 
reuse pretrained DL model in a new and possibly unseen environment.
Our proposed framework can seamlessly integrate with prevailing CSI feedback models, necessitating only the insertion and customization of a translation and retranslation module.
\item We present a plug-in CSI feedback architecture equipped with a lightweight translation module for communication-efficient encoder updates to fit dynamic wireless environments. This module adeptly processes CSI data from new RF environments, aligning it with the format of pretrained data, and ensuring efficient reuse of the pretrained model. By incorporating an effective sparsity aligning function in the angular-delay domain alongside a lightweight CNN, our translation module achieves superior recovery accuracy. Contrary to existing methods that mandate separate encoder networks with millions of parameters, our approach offers a more efficient alternative with only thousands of additional parameters.
 \item In response to the demand for enhanced CSI feedback efficiency in intricate RF environments, we introduce an efficient CSI feedback framework
inspired by CS, named SPTM2-ISTANet+. This model incorporates a spherical feedback structure to regulate input distribution and mitigate path loss effects. We also integrate a deep unfolding decoder network with a residual recovery structure and optimized initialization, complemented by a trainable measurement matrix, to refine the CSI recovery accuracy. Simulation results demonstrate superior performance, including normalized mean square error (NMSE) of $-24.3$ dB 
at compression ratio (CR)
of ${1}/{4}$ in a popular outdoor scenario \cite{ref:csinet}.
\item Instead of employing a blackbox GAN for CSI sample augmentation, we develop a simple but effective model-driven augmentation strategy by exploiting domain knowledge and physical
features of CSI matrices. Taking into consideration
of geographical continuity and delay property of MIMO channels,
we develop a circular shifting 
augmentation of CSI magnitudes in 
angular-delay domain to incorporate the circular 
property of discrete Fourier transform (DFT).  
Additionally, we introduce row-wise random phase variations to CSI elements, simulating the decoupled multipath channel phase shifts. 
Test results demonstrate significantly enhanced CSI recovery performance over blackbox GAN, 
and can achieve NMSE of $-15.8$ dB 
based on only $100$ measurement samples at CR of ${1}/{4}$  in the outdoor scenario.
\end{itemize}}

    
    
    


\section{System Model}

Without loss of generality, we consider a massive MIMO gNB equipped with $N_b$ antennas
to serve a number of single-antenna UEs within its coverage. 
Orthogonal frequency division multiplexing (OFDM) is adopted
in downlink transmission
over $N_f$ subcarriers. 
For subcarrier $m$, 
let $\mathbf{h}_{m} \in \mathbb{C}^{N_b\times1}$ denote the channel vector, 
$\mathbf{w}_{m} \in \mathbb{C}^{N_b\times1}$ denote transmit precoding vector, $x_{m}\in \mathbb{C}$ 
denote the transmitted data symbol, and $n_{m}\in \mathbb{C}$ denote the additive noise. Correspondingly,
the received signal of the UE is
\begin{equation}
	y_{m} =\mathbf{h}_{m}^H\mathbf{w}_{m}x_{m} + n_{m}, 
\label{equ1}
\end{equation}
where $(\cdot)^H$ represents the conjugate transpose.
The downlink CSI matrix in the spatial-frequency domain is denoted by $\tilde{\mathbf{H}} = \left[\mathbf{h}_{1},..., \mathbf{h}_{N_f}\right]^H \in \mathbb{C}^{N_f\times N_b}$.  

To reduce feedback overhead, we first exploit the sparsity of CSI in the delay domain.
Applying 2D DFT, CSI matrix 
$\mathbf{H}_{sf}$ in 
spatial-frequency domain can be transformed  to be $\mathbf{H}_{ad}$ in angular-delay domain using
\begin{equation}
	\mathbf{F}_d^H \mathbf{H}_{sf}  \mathbf{F}_a =  \mathbf{H}_{ad}, \label{idft3}
\end{equation}%
where $\mathbf{F}_d$ and $\mathbf{F}_a $ denote the $N_f \times N_f$ and $N_b \times N_b$ unitary DFT matrices, respectively.  
Owing to limited delay spread and scatters in the practical radio environment, most elements in the $N_f\times N_b$
matrix $\mathbf{H}_{ad}$ are negligibly small except for its
first $R_d$ rows \cite{ref:csinet}. Therefore, we can approximate the MIMO
channel by keeping the first $R_d$ rows
of $\mathbf{H}_{ad}$, denoted by $\mathbf{H}$. 

To allow DL-based CSI feedback system to track the space-varying characteristics of wireless fading channels under different environments, mobile network operators need 
to collect numerous channel measurements in each environment and train a 
corresponding DL network. Furthermore,  UEs entering a new environment 
would need to adjust their encoder DL network which may be large
with millions of parameters. Assuming
$T$ typical scenarios in a region, 
we can use $\mathbf{H}^{t}$
to denote the channel matrix for scenario $t$. 



\textcolor{black}{To diminish deployment overhead and channel measurements of DL models in variational environments for CSI feedback, we introduce a few-shot learning CSI framework that capitalizes on prior training outcomes from recognized channel environments. To leverage prior training outcomes 
from a previously known scenario, 
we design our DL feedback networks for 
CSI feedback under changing scenarios in
two phases: a) constructing a
DL network in an anchor 
scenario without prior 
information ($t = 1$); 
b) constructing DL networks for
additional
scenarios ($t = 2,3,...,T$) 
by leveraging the information learned in prior training phase.}

Let $\mathbf{\hat{H}}^t$ be the recovered CSI matrix 
by the decoder in scenario $t$ corresponding to the ground-truth $\mathbf{{H}}^t$.  
Define the encoding and decoding function as $f_\text{en}(\cdot)$ and $f_\text{de}(\cdot)$, respectively. For downlink CSI feedback architecture in the anchor scenario, the encoder network and decoder network can be denoted, respectively, by 
\begin{eqnarray}
\qquad\qquad	\mathbf{s}_1 &=& f_{\text{en},1}(\mathbf{H}^{1};\Phi_1), 
\\
	\hat{\mathbf{H}}^{1} &=& f_{\text{de},1}(\mathbf{s}_1;\Psi_1).
\end{eqnarray}%
For the subsequent
scenario $t$ ($t\ge 2$), the encoder network and decoder network can be formed, respectively, by 
\begin{eqnarray}
\qquad\qquad	\mathbf{s}_t &=& f_{\text{en},t}(\mathbf{H}^{t};\Phi_t|\Phi_1,\Psi_1), 
\\
	\hat{\mathbf{H}}^{t} &=& f_{\text{de},t}(\mathbf{s}_t;\Psi_t|\Phi_1,\Psi_1).
\end{eqnarray}%


\section{Overall Framework}

\begin{figure}[thpb]
      \centering
      \includegraphics[scale=0.21]{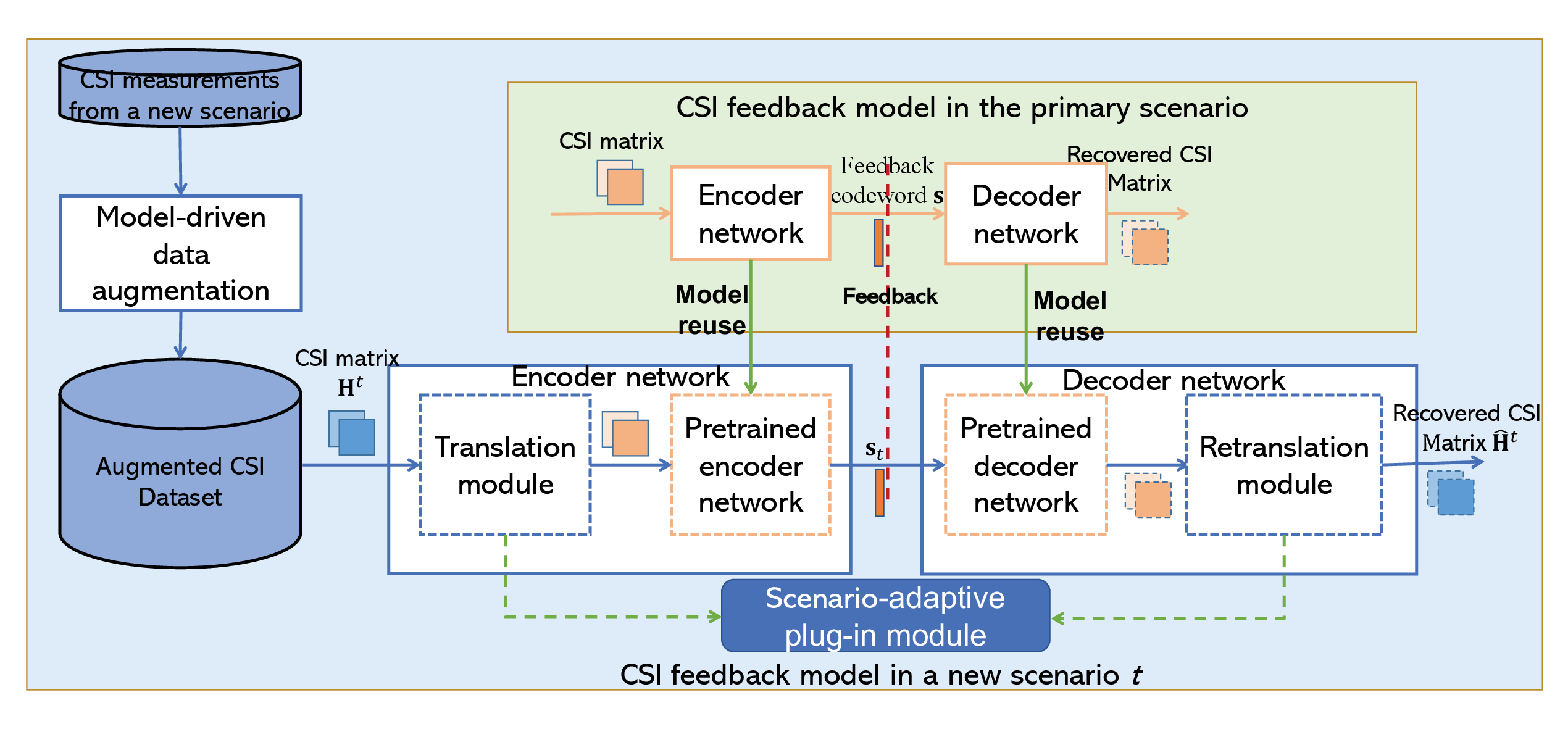}\vspace*{-1mm}
      \caption{Architecture of overall framework}
      \label{figure_overall_framework}
      \vspace*{-2mm}
  \end{figure}

\textcolor{black}{DL networks for CSI feedback are generally optimized via training from
CSI data samples for a customized radio propagation scenario.  To cover a city with 
various scenario types, large numbers of channel measurements are
often required for each likely scenario. Besides, how to update the encoder network when UE is moving among the variational scenarios and sub-areas that exceed the spatial correlation distances is challenging, since a single CSI model tends to exhibit poor performance when applied to other wireless environments due to model mismatch. Given that CSI feedback encoder networks \cite{guo2019convolutional,cui2022,mourya2023} can have millions of parameters, this implies a significant transmission bandwidth overhead. In light of these challenges, this work prioritizes two pivotal objectives:
(1) to design 
a communication-efficient scenario-adaptive DL framework
for accurate CSI recovery;
(2) to develop efficient use of 
limited CSI
data measurement in DL training based on domain
knowledge.}

To achieve these two objectives, we propose a scenario-adaptive CSI feedback framework
that reuses pretrained high-accuracy CSI feedback model 
together with the model-driven data augmentation in 
new environments. As shown in Fig. \ref{figure_overall_framework}, 
the overall framework consists of the following three modules:

\textcolor{black}{
\begin{itemize}
 \item \textbf{Robust Backbone Model for an Anchor Scenario}: We advocate for the establishment of an accurate and robust backbone model suitable for reuse in diverse environments, thereby optimizing deployment costs. Acknowledging that CSI matrices of massive MIMO always exhibit a high degree of variation and suffer from feedback performance degradation as the CSI variation range expands \cite{ref:multitask_csi}, we formulate the SPTM2-ISTANet+ as our backbone network to accommodate complex CSI features and variations. SPTM2-ISTANet+ amalgamates a spherical feedback structure with a CS-inspired residual recovery DNN model, accentuating CSI feedback robustness and precision.  The detailed design of SPTM2-ISTANet+ is introduced in Section IV.
 \item \textbf{Scenario-Adaptive Plug-In Module}: To facilitate the pretrained backbone model's reuse with a lightweight deployment cost, we propose translating CSI matrices in fresh scenarios into a ``style'' compatible with the anchor scenario. Our lightweight plug-in module, designed for scenario adaptation, requires communication-efficient updating for new scenarios, housing merely a few thousand trainable parameters. Integrated translation and retranslation modules facilitate CSI style adjustments and recovery, discussed in detail in Section V.
\item  \textbf{Model-Driven Data Augmentation Module}.  To reduce the cost associated with 
labor-intensive channel measurement, we customize a simple but effective model-driven 
augmentation module by exploiting physical features of CSI matrices based on domain knowledge. 
CSI magnitudes and phases are augmented separately in view of their correlative channel properties. The detailed design of model-driven data augmentation is explained in Section VI.
\end{itemize}
}

\section{SPTM2-ISTANet+}

\textcolor{black}{We now construct an efficient deep unfolding network as the backbone model for the anchor scenario to improve CSI feedback accuracy. Note that the deep unfolding modules are only used in the decoder network.}

\begin{figure*}[thpb]
      \centering
      \includegraphics[scale=0.31]{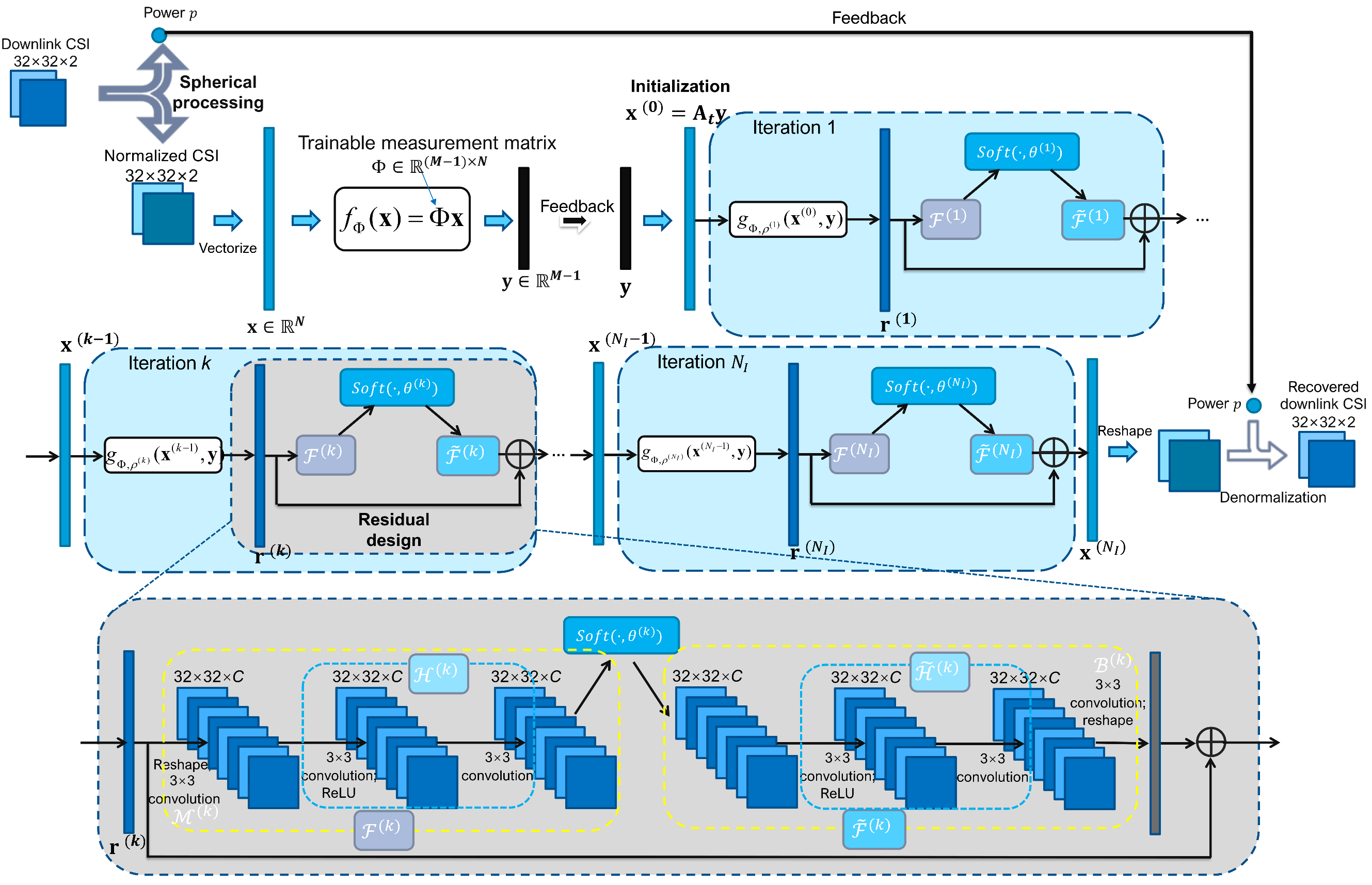}\vspace*{-1mm}
      \caption{Architecture of SPTM2-ISTANet+.}
      \label{figure_architecture}
      \vspace*{-5mm}
  \end{figure*}

\vspace*{-2mm}
\subsection{Encoding Network}
 \vspace*{-1mm}

To improve the robustness and CSI compression efficiency, we adopt a deep 
unfolding-based feedback network, and propose two key innovations to the 
encoder network in view of physical CSI features.

First, we construct a spherical CSI 
feedback structure in view of domain-specific characteristics of the wireless channel. 
The distribution of MIMO CSI coefficients is
substantially different from that of image data.
The dynamic range of CSI is always much greater because of 
radio path loss, as CSI of one UE may differ from that of another
by orders of magnitude.
A naive processing can render CSI of some
UE too small, leading to large recovery errors. 
Hence, before applying the CS measurement matrix to lower
CSI dimensionality, we split CSI matrix $\mathbf{H}$ into a power value $p$ and a 
spherical matrix $\check{\mathbf{H}}$, where 
$p = \Arrowvert\mathbf{H}\Arrowvert$ is the power and $\check{\mathbf{H}} = \mathbf{H}/\Arrowvert\mathbf{H}\Arrowvert$
is a unit norm spherical CSI.

As shown in Fig. \ref{figure_architecture}, after the spherical processing,  we vectorize the normalized downlink CSI matrix $\check{\mathbf{H}}$ as the input to DNN for compression. The real part and imaginary part are split for easier processing. The corresponding vector is denoted as $\mathbf{x} \in \mathbb{R}^{N}$, where $N = 2 \times R_d \times N_b$. To shorten $\mathbf{x}$, a
measurement matrix $\Phi \in \mathbb{R}^{(M-1) \times N}$ is used for dimension compression, where $M = \text{CR}\times N - 1$. 

Second, to deliver further performance improvement, we move away from the traditional 
random constructed measurement matrix in CS and devise a data-driven 
trainable measurement matrix $\Phi$.  The goal is to better capture 
features of massive MIMO for CSI encoding, particularly if the compression degree 
is high (i.e. CR is small). Using only a matrix multiplication at
the encoder, UE computation cost is modest.

\vspace*{-1mm}
\subsection{Decoding Network}
\vspace*{-1mm}

Assuming lossless feedback \cite{ref:csinet,dl_multires}, the low-dimension vector $\mathbf{y} \in \mathbb{R}^{M-1}$ received by gNB can be defined as $\mathbf{y}=\Phi\mathbf{x}$. From $\mathbf{y}$, the decoder network reconstructs the original $\mathbf{x}$ by solving the following compressive sensing  recovery problem:
\begin{equation} \label{eq:ista}
    \min _{\mathbf{x}} \frac{1}{2}\|\Phi \mathbf{x}-\mathbf{y}\|^{2}+\lambda\|\mathcal{F}(\mathbf{x})\|_{1},
\end{equation}
where  $\lambda$ is the regularization parameter, $\|\cdot\|$ denotes the $l_2-$norm. $\mathcal{F}(\cdot)$ is the sparse transform function of $\mathbf{x}$.

Our decoder network of SPTM2-ISTANet+ utilizes the deep unfold structure. Adopting the settings of ISTANet+ \cite{istanet} to unfold the iterative shrinkage-thresholding algorithm (ISTA) \cite{beck2009fast}, we 
recover CSI
by iterating between two steps:
\begin{equation} \label{eq:r}
    \mathbf{r}^{(k)}=\mathbf{x}^{(k-1)}-\rho \Phi^{\top}\left(\Phi\mathbf{x}^{(k-1)}-\mathbf{y}\right),
\end{equation}
\vspace*{-5mm}
\begin{equation} \label{eq:x}
    \mathbf{x}^{(k)}=\underset{\mathbf{x}}{\arg \min } \frac{1}{2}\left\|\mathbf{x}-\mathbf{r}^{(k)}\right\|^{2}+\lambda\|\mathcal{F}(\mathbf{x})\|_{1},
\end{equation}
where $k$ denotes the iteration index, $\rho$ denotes the step size. Next, 
we expand Eq.~(\ref{eq:r}) and Eq.~(\ref{eq:x}) respectively into deep unfolding modules corresponding to the $k$-th iteration, as model $\mathbf{r}^{ (k)}$  and module $\mathbf{x}^{(k)}$, to solve the recovery problem.



Module $\mathbf{r}^{(k)}$ corresponds to Eq.~(\ref{eq:r}) and generates $\mathbf{r}^{(k)}$ from the result of the $(k-1)$-th iteration. In order to improve flexibility of the recovery network, step size $\rho$ in Eq.~(\ref{eq:r}) may be automatically adjusted according to iteration, i.e., $\rho^{(k)}$ varies for each $k$. Therefore, module $\mathbf{r}^{(k)}$ can be viewed as a function of $\mathbf{x}^{(k-1)}$ and $\mathbf{y}$, i.e.,
\begin{equation} \label{eq:rk}
    \mathbf{r}^{(k)}\hspace*{-2mm}=g_{\Phi,\rho^{(k)}}
    \hspace*{-1mm}\left(\mathbf{x}^{(k-1)}\hspace*{-2mm}, \mathbf{y}\right)=\mathbf{x}^{(k-1)}\hspace*{-1mm}-\rho^{(k)} \Phi^{\top}\hspace*{-1mm}\left(\Phi\mathbf{x}^{(k-1)}-\mathbf{y}\right).
\end{equation}

Module $\mathbf{x}^{(k)}$ corresponds to Eq. (\ref{eq:x}) and calculates $\mathbf{x}^{(k)}$ from $\mathbf{r}^{(k)}$ in the $k$-th iteration. 
A combination of two convolutional layers and a ReLU unit (i.e. a diode) $\text{ReLU}(x)=\max(0, x)$ is used to construct the sparse transformation $\mathcal{F}(\cdot )$ in Eq.~(\ref{eq:x}), i.e., $\mathcal{F}(\mathbf{x})=\mathbf{B} \cdot \text{ReLU}(\mathbf{A} \mathbf{x})$.
Both $\mathbf{A}$ and $\mathbf{B}$ use convolutional layers without bias to achieve equivalent matrix operations. To overcome the vanishing gradient issue which often leads to poorer performance in deep unfolding,  a residual structure is constructed 
 to enhance recovery accuracy. 

From Eq.~(\ref{eq:x}), we assume that $\mathbf{x}^{(k)}=\mathbf{r}^{(k)}+\mathbf{w}^{(k)}+ \mathbf{e}^{(k)}$, where $\mathbf{w}^{(k)}$ represents the missing high-frequency components in $\mathbf{r}^{(k)}$, and $\mathbf{e}^{(k)}$ denotes noise. We then apply a linear operation $\mathcal{R}(\cdot)$ to extract missing component $\mathbf{w}^{(k)}$ from $\mathbf{x}^{(k)}$, i.e., $ \mathbf{w}^{(k)}=\mathcal{R}\left(\mathbf{x}^{(k)}\right)$. Define $\mathcal{R}(\cdot)$ as $\mathcal{R}=\mathcal{B}\circ\mathcal{M}$, where $\mathcal{M}$ and $\mathcal{B}$ corresponds to a convolutional layer without bias terms with
kernel size $3\times3$. 
Note that when a sparse transformation satisfies $\mathcal{F}(\mathbf{x})=\mathbf{B} \text{ReLU}(\mathbf{A} \mathbf{x })$, the following approximation holds:
$\left\|\mathcal{F}(\mathbf{x})-\mathcal{F}\left(\mathbf{r}^{(k)}\right)\right\|^{2} \approx \alpha\left\|\mathbf{x}-\mathbf{r}^{(k)}\right\|^{2}$  \cite{istanet},
where $\alpha$ is a scalar only related to parameters of transformation $\mathcal{F}(\cdot)$.
Next, decompose $\mathcal{F}^{(k)}$ into $\mathcal{F}^{(k)}=\mathcal{H}^{(k)}\circ\mathcal{M}^ {(k)}$, where $\mathcal{H}^{(k)}$ consists of two convolutional layers without bias plus a ReLU activation. 
Eq.~(\ref{eq:x}) can be transformed into
\begin{align} \label{eq:xk2}
    \mathbf{x}^{(k)}=\underset{\mathbf{x}}{\argmin } &\frac{1}{2}\left\|\mathcal{H}^{(k)}(\mathcal{M}^{(k)}(\mathbf{x}))-\mathcal{H}^{(k)}\left(\mathcal{M}^{(k)}\left(\mathbf{r}^{(k)}\right)\right)\right\|^{2} \notag \\ 
    &+\theta^{(k)}\|\mathcal{H}^{(k)}(\mathcal{M}^{(k)}(\mathbf{x}))\|_{1}.
\end{align}
Next, construct the left inverse function of $\mathcal{H}^{(k)}(\cdot)$ such that $\widetilde{\mathcal{H}}^{(k)} \circ \mathcal{H}^ {(k)}=\mathcal{I}$, where $\mathcal{I}$ is the identity operator. We then can use a DNN to construct a symmetric structure of $\widetilde{\mathcal{H}}^{(k)}(\cdot)$ as $\mathcal{H}^{(k)}(\cdot)$, and include
the constraint of $\widetilde{\mathcal{H}}^{(k)} \circ \mathcal{H}^{(k)}=\mathcal{I}$ to the loss function. Finally, a closed-form 
expression of $\mathbf{x}^{(k)}$ can be
\begin{equation} \label{eq:xk3}\hspace*{-2mm}
\mathbf{x}^{(k)}\hspace*{-1mm}=\hspace*{-1mm}\mathbf{r}^{(k)}\hspace*{-1mm}+\mathcal{B}^{(k)}\hspace*{-1mm}\left[\widetilde{\mathcal{H}}^{(k)}\hspace*{-1mm}\left[
\mbox{soft}\hspace*{-1mm}\left[\mathcal{H}^{(k)}\hspace*{-1mm}\left(\mathcal{M}^{(k)}\hspace*{-1mm}\left(\mathbf{r}^{(k)}\hspace*{-1mm}\right)\hspace*{-1mm}\right), \theta^{(k)}\right]\right]\right],
\end{equation}
where we define a soft threshold function 
$\mbox{soft}(x, \theta)=\operatorname{sgn}(x) \max (0, |x|-\theta)$.
The network structure corresponding to module $\mathbf{x}^{(k)}$
is shown in the gray box at Fig. \ref{figure_architecture}.
Kernel number $C$ is set to $32$ by default.  \textcolor{black}{In summary, the trainable parameters in SPTM2-ISTANet+ are $\left\{\rho^{(k)}, \theta^{(k)},\mathcal{H}^{(k)}, \widetilde{\mathcal{H}}^{(k)}, \mathcal{M}^{(k)}, \mathcal{B}^{(k)}, \Phi\right\}$.}


\textcolor{black}{To reconstruct the pre-compressed CSI vector \( \mathbf{x} \) over \( N_I \) iterative modules, an initial estimate \( \mathbf{x}^{(0)} \) is essential. Instead of initializing with zero values as done in \cite{wang2020}, the least squares estimation is utilized for initialization, i.e., $\mathbf{x}^{(0)} = \mathbf{A}_{t}\mathbf{y}$, 
where \( \mathbf{A}_{t} \) represents the least square estimate of the transformation matrix \( \mathbf{A} \) in the linear transformation \( \mathbf{x} = \mathbf{A} \mathbf{y} \). It is computed as \( \mathbf{A}_{t} = \mathbf{XY}^{\top} (\mathbf{Y} \mathbf{Y}^{\top})^{-1} \),
with \( \mathbf{X} \in \mathbb{R}^{N \times N_T} \) and \( \mathbf{Y} \in \mathbb{R}^{M \times N_T} \) being matrices formed from the column vectors \( \mathbf{x} \) and \( \mathbf{y} \) of the training set, respectively. Here, \( N_T \) denotes the total number of training samples.}


To optimize parameters in SPTM2-ISTANet+, 
we need an efficient loss function for training. 
Define the size and the $n$-th CSI vector of the training set, respectively,
as $N_T$ and $\mathbf{x}_n \in \mathbb{R}^{N}$. Define the number of iteration modules as 
$N_I$. We can construct loss function
\begin{equation}
\mathcal{L}_{\text {total}}(\Theta)=\mathcal{L}_{\text {MSE}}+\gamma\cdot \mathcal{L}_{\text {constraint}},
\end{equation}
where the mean square error (MSE) $\mathcal{L}_{\text {MSE}}=\frac{1}{N_{T} N} \sum_{n=1}^{N_{T}}\left\|\mathbf{ x}_{n}^{\left(N_{I}\right)}-\mathbf{x}_{n}\right\|^{2}$ is the CSI reconstruction accuracy indicator, which is commonly used in the CSI feedback. Additionally,  
$\mathcal{L}_{\text {constraint}} = \frac{1}{N_{T}N} $ $ \sum_{n=1}^{N_{T}} \sum_{k= 1}^{N_{I}}\left\|\widetilde{\mathcal{H}}^{(k)}\left(\mathcal{H}^{(k)}\left(\mathcal{M} ^{(k)}\left(\mathbf{r}_n^{(k)}\right)\right)\right) - \mathcal{M}^{(k)} \\ \left(\mathbf{r} _n^{(k)}\right)\right\|^{2}$ corresponds to $\widetilde{\mathcal{H}}^{(k)} \circ \mathcal{H}^{(k)}= \mathcal{I}$ restriction, with which $\gamma$ is the regularization weight
(set to $0.01$ unless noted otherwise).

\section{CSI Translation Architecture CSI-TransNet}
\begin{figure}[thpb]
      \centering
      \includegraphics[scale=0.24]{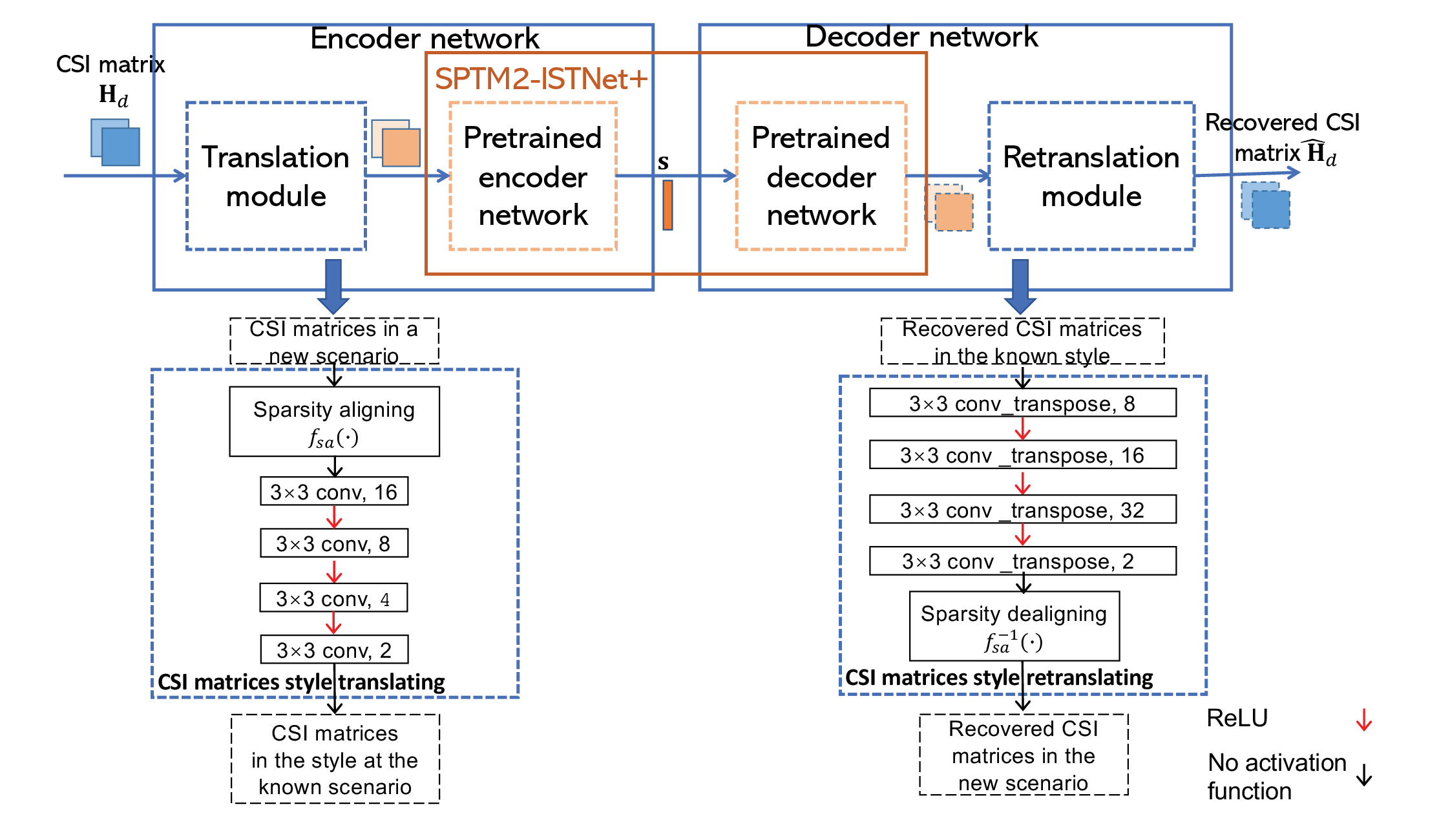}\vspace*{-1mm}
      \caption{Architecture of CSI-TransNet.}
      \label{figure_transnet}
      \vspace*{-5mm}
  \end{figure}

Tackling CSIs from various environments is one of the key challenges in
the practical deployment of DL-based CSI feedback schemes. One solution is to train a 
customized network for each scenario or region, and dynamically switch to the 
feedback network according to UE-detected channel environments. 
However, this solution requires UE to have sufficient memory to save 
many DL-based encoder networks, each of which may include millions of parameters. 
Alternatively, UE may frequently update its encoder by downloading new model parameters, 
thereby consuming high cost in terms of wireless bandwidth and UE energy. 

Image-to-image translation \cite{ref:imag_trans} has been used in computer vision to handle a variety of problems including image stylization \cite{ref:style_trans}
and segmentation \cite{ref:segmentation}.  It aims to learn a mapping that can convert an image from a source domain to a target domain, while preserving the main presentations of input images. For example, a horse image can be converted to an image similar to the zebra’s style. 
Motivated by the application of image-to-image translation, we propose
to translate the style of CSI in a new scenario to match the style in the
known scenario of a given CSI feedback model. In this section, we propose an efficient CSI feedback architecture ``CSI-TransNet'', which incorporates
a lightweight module on the UE side to overcome performance obstacles of DL-based CSI model in various environments. 
By exploiting the CSI-to-CSI translation, CSI-TransNet can reuse the 
pretrained CSI model with high recovery accuracy in a new environment. 

The full CSI-TransNet architecture is shown in Fig. \ref{figure_transnet}, where the
encoder network at UE is equipped with one translation module and one shared pretrained 
encoder network. The decoder network at gNB contains a shared pretrained decoder network 
and one 
customized retranslation module. Note that, the translation and retranslation modules are 
two plug-in 
modules that only need a few thousand
parameters that can be easily updated each time when
a UE encounters a new channel environment.

CSI-to-CSI translation makes it easier for CSI matrices after translation 
to be efficiently compressed and accurately recovered by the pretrained model 
without further tuning. Consequently, CSI matrices after translation should 
have a similar property to that of CSI matrices in the anchor scenario. 
Unlike data-based translation, as shown in Fig. \ref{figure_transnet},  
we customize an effective sparsity aligning function before
the DL-based translation network in the translation module to 
achieve lightweight simplicity.

Since the CSI matrices are sparse in angular-delay domain \cite{ref:csinet}, we apply
circular shift in angular-delay domain to CSI matrices in a new scenario to 
promote a similar sparsity to the anchor scenario used to pre-train
the anchor model. Define the sparsity aligning function as $f_\text{sa}(\cdot)$, and define the circular shift function as $f_\text{sh}(\cdot,i,j)$, where $i$ and $j$ are 
shift steps in row and column, respectively. For a given CSI matrix $\mathbf{H}$,
elements of CSI matrix after the circular shift $\mathbf{H}^\text{sh}=f_\text{sh}(\mathbf{H},i,j)$ can be written as 
\begin{equation}
\mathbf{H}^\text{sh}_{m,n}=\mathbf{H}_{(m-i)\bmod R_d,(n-j)\bmod N_b},\forall m\in\left[R_{d}\right], n \in\left[N_{b}\right]. 
\end{equation}
Consequently, the goal of the sparsity aligning function is to find shift steps $i$ and $j$ to achieve the best similarity. One way to achieve this goal 
is to calculate the shift step corresponding to the best circular cross-correlation \cite{ref:wang2019kernel} for CSI magnitude matrices from
the two scenarios. A more strict and convenient way is to calculate the shifting steps
$i$ and $j$ corresponding to the best CSI recovery accuracy of the 
pretrained CSI feedback network, i.e.,
\begin{equation}
    \min_{i,j} 
    \sum_{n}\left \|f_\text{sh}(\mathbf{H}_n,i,j)-f_{\text{de},1}( f_{\text{en},1}(f_\text{sh}(\mathbf{H}_n,i,j);\Phi_1);\Psi_1)\right \|^2
\label{second}
\end{equation}
where $\mathbf{H}_n$ corresponds to the $n$-th measured CSI matrix in the new channel
scenario. Owing to the translation invariance of the convolutional layers, we 
propose to use the second method of (\ref{second}) to determine the 
shift steps $i$ and $j$. Given a pair of selected shift steps $i$ and $j$, we can get the CSI matrix after sparsity aligning in the scenario $t$ as
\begin{equation}
\mathbf{H}^\text{sa}=f_\text{sa}(\mathbf{H}^{t})=f_\text{sh}(\mathbf{H}^t,i,j).
\end{equation}

After sparsity aligning, DNN of the translation module consists of $4$ convolutional layers for feature extraction and translation. Specifically, $4$ convolutional layers
utilize $3\times 3$ kernel to generate $16$, $8$, $4$ and $2$ feature maps, respectively. The first $3$ convolutional layers use the ReLU activation function.

The retranslation module mirrors the structure of the translation module, where a DL network is constructed first to fine-tune the CSI matrix, followed by a sparsity dealigning function $f_\text{sa}^{-1}(\cdot)$ to restore sparsity in the new scenario.
To reverse convolution, transposed convolution layers \cite{ref:conv_tranposed} are used in the retranslation module. Specifically,  $4$ transposed convolutional layers
utilize $3\times 3$ kernel to generate $32$, $16$, $8$ and $2$ feature maps, respectively. 
The first $3$ transposed convolutional layers use
ReLU activation function. Using the same shifts steps $i$ and $j$ from
sparsity aligning, we recover the CSI matrix in the scenario $t$ as

 \begin{equation}
\hat{\mathbf{H}}^{t}=f_\text{sa}^{-1}(\hat{\mathbf{H}}^\text{sa})=f_\text{sh}(\hat{\mathbf{H}}^\text{sa},-i,-j).
\end{equation}

Let $\Theta_t$ and $\Omega_t$ be DNN parameters in the translation module and retranslation module, respectively. We write
translation function $f_\text{tra}(\cdot; \Theta_t)$ and retranslation function
$f_\text{ret}(\cdot;\Omega_t )$. For CSI-TransNet in the scenario $t$, after determining the shift steps $i$ and $j$, we jointly train only parameters in the plug-in modules $f_\text{tra}(\cdot; \Theta_t)$ and $f_\text{ret}(\cdot;\Omega_t)$ using the following loss function $    \mathcal{L}_{\text {MSE}}$:
\begin{equation*}
\frac{1}{N_t}\Sigma_{n}\left \|\mathbf{H}^\text{sa}_n-\hspace*{-1mm}f_{\text{det},t}(f_{\text{de},1}\hspace*{-0.5mm}( f_{\text{en},1}\hspace*{-0.5mm}(f_{\text{tra},t}\hspace*{-0.5mm}(\mathbf{H}^\text{sa}_n; \Theta_t);\Phi_1);\Psi_1);\Omega_t)\right \|^2,
\end{equation*}
where $N_t$ is the number of training CSI data samples for the $t$-th scenario.


When a UE detects and reports a new scenario, the gNB serving this area in the CSI-TransNet architecture sends weights of the translation module to the UE to ensure 
CSI feedback accuracy. In view of the lightweight design, only a few thousand parameters are included, which significantly lightens the payload in comparison with the
total number of encoder parameters.





\section{Model-driven Data Augmentation}

\begin{figure}[thpb]
      \centering
      \includegraphics[scale=0.24]{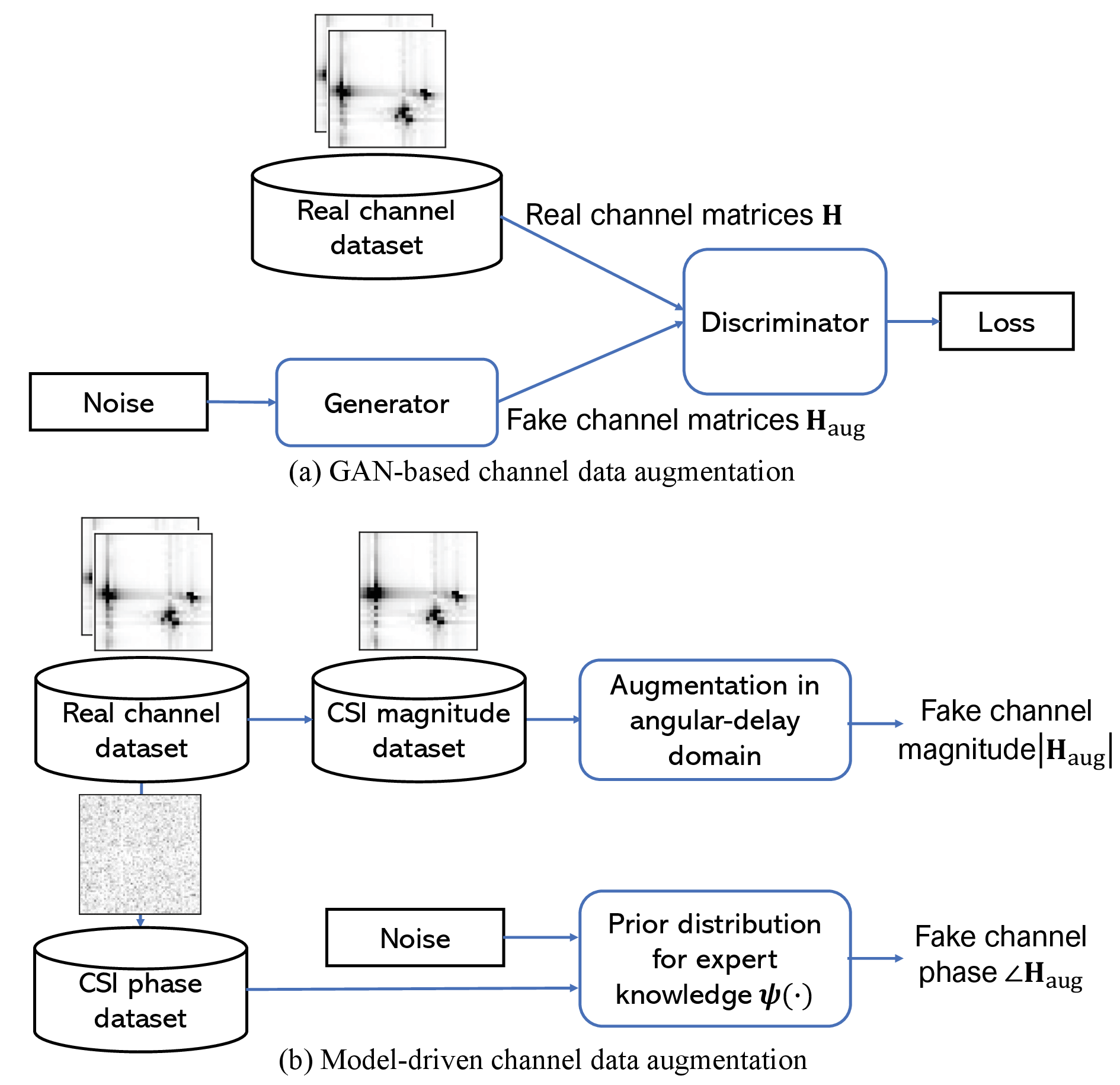}
      \caption{Architecture comparison between proposed data augmentation and GAN-base data augmentation.}
      \label{figure_gan_org} \vspace*{-3mm}
  \end{figure}
In addition to lowering the deployment cost of CSI feedback models in variational environments, 
we also consider the challenge of training DL models for CSI feedback which uses at least
tens of thousands of real-time channel measurements
in each scenario. Many DL models have adopted data
augmentation as an effective solution.  We investigate
how to leverage domain knowledge for data augmentation
in compressive CSI feedback and recovery. 

\subsection{GAN-based Augmentation}
Fig. \ref{figure_gan_org}(a) shows a traditional GAN-based approach to data augmentation that alternately solves maximizing and minimizing optimization problems during training to reduce the distance between the distributions of generated channels and real channels. The generator uses a DNN to map the Gaussian noise vector 
to generate imitations (fake channels). When GAN  converges to generate channels with matching distribution to true channels, we can treat the GAN
as a stable storage of CSI models for providing 
augmented channel data to form a large training dataset to support CSI-TransNet training. However, GAN itself
requires sufficient data samples to train. Furthermore, when available CSI measurement samples only 
partially represent
RF channel features in a coverage area, even a
well-trained generator can be a heavily biased channel model which can severely degrade the performance 
of compressive CSI models trained with biased CSI data. 
In short,  it is not enough to obtain sufficient data samples in GAN-based training. It is vital for 
augmented samples to cover features that are absent
from existing measurement samples.


\subsection{Model-driven Augmentation}

To enhance DNN training for CSI feedback, we note that augmented samples should present absent or under-represented features among existing measurements.
Leveraging domain knowledge, we develop a simple but effective model-driven augmentation by decoupling the characteristics in the magnitude and phase of CSI matrices. 

We begin by first split phase and magnitude of MIMO channel matrices before augmentation:
\begin{equation}
\mathbf{H}=|\mathbf{H}| \odot e^{j \angle \mathbf{H}},
\end{equation}
where $\odot$ denotes Hadamard product. The $(m, n)$-th entry of $\mathbf{H}$ is written as $\mathbf{H}_{m, n}=\left|\mathbf{H}_{m, n}\right| e^{j \angle \mathbf{H}_{m, n}}$. This way, the magnitude CSI matrix is $|\mathbf{H}|$ with entries $\left|\mathbf{H}_{m, n}\right|$ and the phase matrix is
written as $\angle \mathbf{H}$ with entries $\angle \mathbf{H}_{m, n}$. 
This split allows us to apply domain knowledge about CSI features including multipath delay profile and phase distribution.

Next, we utilize the geographical continuity of 
CSI variation to generate augmented magnitude matrices, which should exhibit similar characteristics to
measured channels. Given a typical environment with
fixed paths between gNB and UE, it has been noted that geographically continuous UE movement should lead to smooth variation in the angular-delay domain \cite{dyloc}. In other words, CSIs in the vicinity of a measurement spot are highly correlated in the angular-delay domain because of similar  arrival/departure angles and delays of
multipath propagation. Consequently, we can construct multiple angle-delay profiles by (circularly) 
shifting the CSI magnitude matrix in angular-delay domain to generate
new CSI matrices that reflect features of nearby UE CSIs.
Leveraging the circular characteristic 
of CSI matrices in the angular-delay domain 
based on the property of DFT, the shifts are 
circular. In short, entries of the augmented magnitude
CSI matrix $\left|\mathbf{H}^\text{aug}_{m, n}\right|$  are generated by the following 
rule in the angular-delay domain, 
\begin{equation}
    \left|\mathbf{H}^\text{aug}_{m, n}\right|=\left\{\begin{matrix}\left|\mathbf{H}_{m+i,(n+j)\bmod N_b}\right|,&
 1\le m+i\le R_d,\\
0\qquad, &\text{else}, 
\end{matrix}\right. 
\end{equation}
$\forall m\in\left[R_{d}\right], n \in\left[N_{b}\right]$. Note that we
select shift steps $\lfloor -\frac{R_{d}}{2}\rfloor \le i \le  \lfloor \frac{R_{d}}{2}\rfloor$ and $\lfloor -\frac{N_{b}}{2}\rfloor \le j \le  \lfloor \frac{N_{b}}{2}\rfloor$ in angular and delay domain, respectively. We apply truncation in the delay domain by setting to zero delay elements beyond $R_d$ rows.

We also augment the phase matrices.
The augmented phases should cover cases beyond the measured CSIs to enhance training and avoid overfitting. Considering that the path to antenna array arrived at the same delay can share frequency-independent phase shift that includes direct or reflected path \cite{tse2005fundamentals}, we select uniform distribution as augmented phase shifting distribution  $\boldsymbol{\psi}_\text{aug}(\cdot) \sim \mathcal{U}(0, 2\pi)$ to elements in the same row (i.e., same delay).
In other words, we construct a larger phase variation than the measured CSI phase, and we
use recall to replace some precision to enhance 
 CSI recovery accuracy in the practical
deployment.
Accordingly, for each 
row of a sampled CSI matrix, 
we apply a random phase shift:
\begin{equation}
 \angle\mathbf{H}^{\text {aug}}_{m, n}= \angle\mathbf{H}_{m, n}+\angle e^{-j \theta_{m}} , \forall m \in\left[R_{d}\right], n \in\left[N_{b}\right],
\end{equation}
where $\theta_{m} \sim \mathcal{U}(0, 2\pi)$.


Finally, we combine augmented magnitude and phase matrices to generate CSI samples for 
training CSI feedback models.



\vspace*{-1mm}
\section{Performance Evaluation}

\subsection{Experiment Setup}
\vspace*{-1mm}

For performance evaluation, we use the following four datasets, where the first two dataset settings are typically used for assessing the performance of CSI estimation and feedback techniques \cite{ref:csinet}, the third dataset is from actual field measurements, and the last dataset is generated following 3GPP standard models:
\begin{enumerate}
\item \textbf{Cost2100 Indoor}. The dataset is generated  by using COST2100 model \cite{c2100} with $5.3$ GHz downlink frequency for a gNB at the center of a $20$m$\times 20$m coverage area. The bandwidth is $20$ MHz. We consider $N_b=32$ antennas and $N_f = 1024$ subcarriers at the gNB to
serve single-antenna UEs randomly distributed within
the coverage area. 
\item \textbf{Cost2100 Outdoor}. The dataset is generated from COST2100 model \cite{c2100} for a $300$ MHz downlink frequency for a gNB at the center 
of a $400$m$\times 400$m coverage area. The bandwidth, antennas and subcarriers are the same as in 
Cost2100 Indoor.

\item \textbf{Measured Indoor}. This dataset contains CSI samples recorded from KU Leuven Massive MIMO testbed in a 9 $\mbox{m}^2$ indoor area \cite{ref:data_ula}. The gNB is equipped with a uniform linear array of 64 antennas at a $2.61$ GHz frequency. We select the first 32 of 64 antennas. 

\item \textbf{Quadriga 3GPP UMA}.  We generate one more dataset using the channel setting
described in 3GPP TR 38.901 with the QuaDRiGa platform \cite{ref:data_quadriga}.  We select the urban macrocell (UMa) scenario at the $2.6$ GHz carrier over
$20$ MHz bandwidth with $N_b=32$ antennas. 
The gNB is located at the center of a square area of edge
length $400$m and serves single-antenna UEs randomly distributed within
the coverage area.
\end{enumerate}
After transforming CSI matrices into the angular-delay domain, 
only the first 32 rows are kept owing to sparsity for these four types of channel data. 
The overall training set size is $100,000$ and the testing set size is $20,000$.  The batch size is $64$. We use 200 epochs for SPTM2-ISTANet+, and 80 epochs for the translation module and retranslation module in CSI-TransNet because they have very few parameters. 

\textcolor{black}{To compare the recovery accuracy of different networks, we adopt the metric of normalized MSE, i.e., 
$\textrm{NMSE} = \frac{1}{N_k}
\sum_{k=1}^{N_k}\Arrowvert\mathbf{H}_k-\mathbf{\hat{H}}_k\Arrowvert^2/\Arrowvert\mathbf{H}_k\Arrowvert^2,
$
where $\mathbf{\hat{H}}$ denotes the recovered \( \mathbf{H} \), and \( k \) and \( N_k \) represent the sample index and the total sample count in the test set, respectively. In addition to NMSE, we also measure the cosine similarity to gauge precoding performance.  The cosine similarity metric, \( \rho \), is expressed as:
\begin{equation}
    \rho=\mathbb{E}\left\{\frac{1}{N_{\mathrm{c}}} \sum_{m=1}^{N_{\mathrm{c}}} \frac{\left|\hat{\mathbf{h}}_m^H \mathbf{h}_m\right|}{\left\|\hat{\mathbf{h}}_m\right\|_2\left\|\mathbf{h}_m\right\|_2}\right\},
\end{equation}
where \( \hat{\mathbf{h}}_m \) signifies the reconstructed channel vector for the \( m \)-th subcarrier. When the gNB uses $\mathbf{v}_{ m}=\hat{\mathbf{h}}_{m} /\left\|\hat{\mathbf{h}}_{m}\right\|$ as a beamforming vector (i.e., as in zero-forcing precoding), the cosine similarity can be indicative of the precoding gain.}


\subsection{Feedback Accuracy Comparison of the Anchor Model}

\textcolor{black}{We compare SPTM2-ISTANet+ with six known schemes 
with demonstrated performance in massive MIMO system CSI feedback (without relying on additional auxiliary information, e.g., uplink CSI, previous CSIs): 
\begin{itemize}
    \item CsiNet+ \cite{guo2019convolutional}:
A CSI feedback model that uses larger convolution kernels $(7\times7)$ and optimizes the structure of residual units. 
    \item DCRNet \cite{tang2021dilated}: A CSI feedback model  that combines multiple resolution convolution kernels and dilated convolutions to extract CSI features of the different granularities. It optimizes learning rate adjustment with the help of a warm-up process.
    \item ISTANet+: A design inspired by compressive image processing \cite{istanet} for CSI feedback in \cite{ref:mason} which adopts orthogonal random Gaussian measurement matrix
    $\Phi \in \mathbb{R}^{M \times N}$, where $N = 2048$, $M$ depends on CR. 
     \item TiLISTA-Joint \cite{wang2020}: Another compressive sensing-inspired CSI feedback network with a learnable sparse transformation module to enhance the CSI recovery accuracy.  Different from TiLISTA-Joint, SPTM2-ISTANet+ exploits the spherical structure to mitigate the effect of the channel variation, and residual recovery design and the least squares estimation for initialization instead of initializing with zero values are utilized to enhance the recovery accuracy and convergence. We use the 9-iteration TiLISTA-Joint results in \cite{wang2020} for the Cost 2100 indoor scenario, and we modify the itemized key differences in SPTM2-ISTANet+ to mimic the  TiLISTA-Joint (termed as simulative TiLISTA-Joint) for the Cost 2100 outdoor scenario. 
     \item TransNet \cite{cui2022} and STNet \cite{mourya2023}: TransNet introduces an attention mechanism by a two-layer Transformer architecture, enabling CSI to learn the connections between its parts in the process of feedback. STNet is a lightweight transformer-based model that uses a spatially separable attention mechanism with less complexity.
\end{itemize}}


  \begin{figure}[!htb] 
  \hspace*{-3mm}
	\subfigure[Cost2100 Indoor] {\label{fig:sphista_indoor} 
	\includegraphics[width=0.25\textwidth]{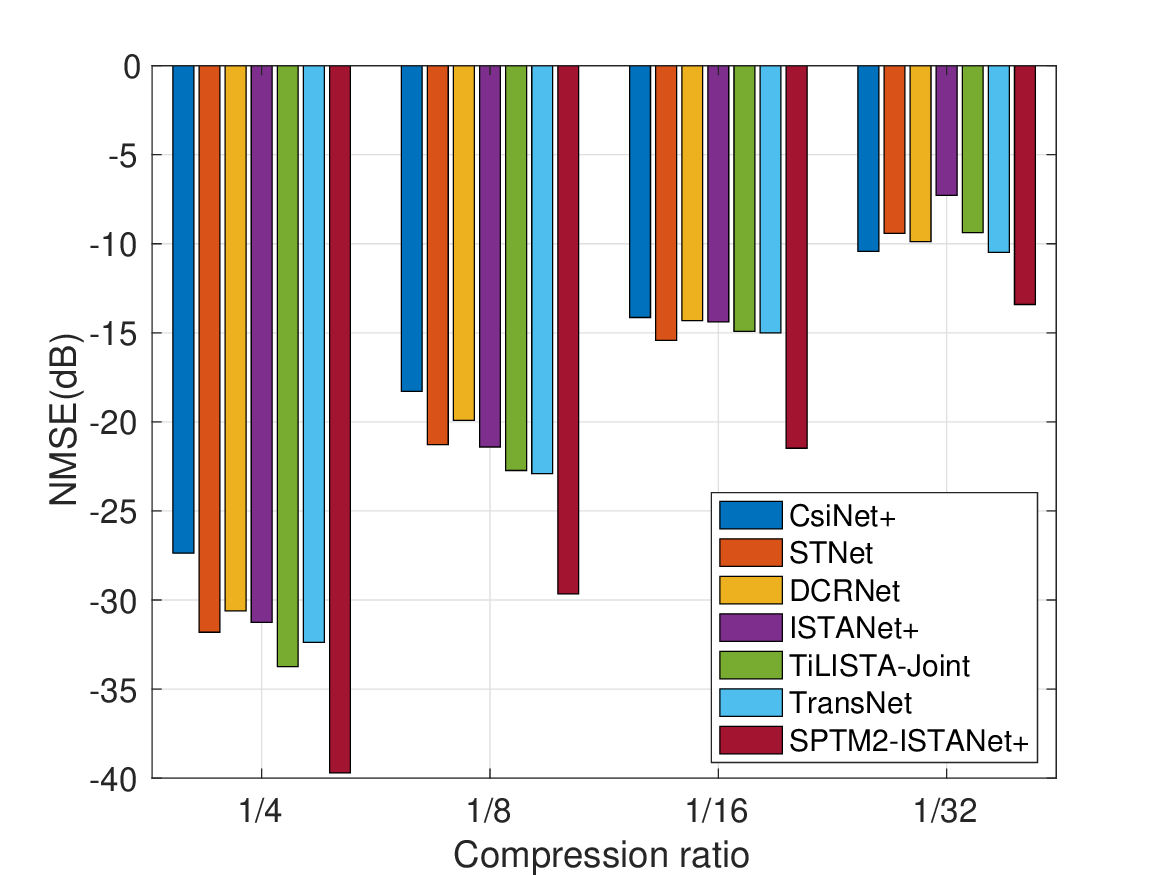}
	} \hspace*{-8mm}
	\subfigure[Cost2100 Outdoor] { \label{fig:sphista_outdoor} 
	\includegraphics[width=0.25\textwidth]{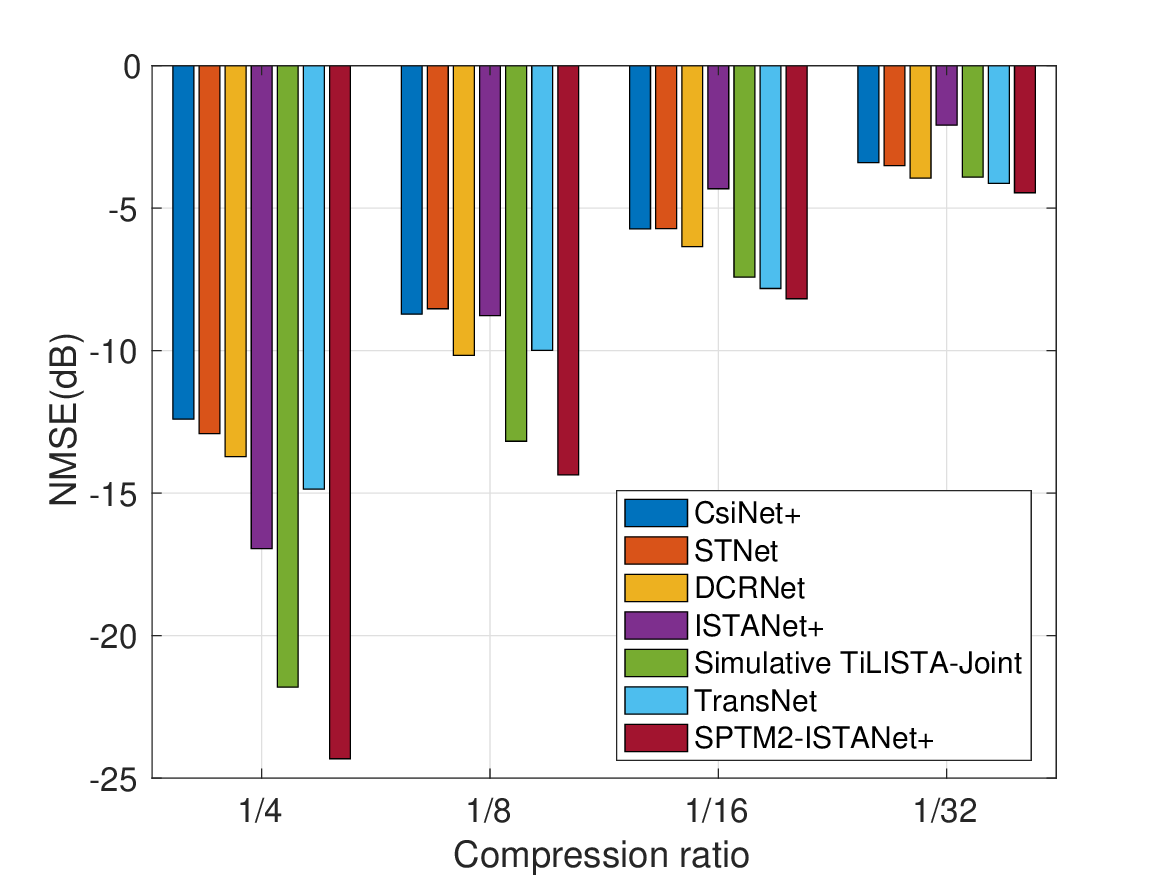} 
	} 
	\hspace*{-3mm}
        \caption{NMSE comparison in different CRs.} 
	\label{fig:nmse_init} 
\end{figure}


\textcolor{black}{Fig. \ref{fig:nmse_init} shows the CSI performance comparison among the seven schemes CsiNet+, STNet, DCRNet, ISTANet+, TiLISTA-Joint, TransNet and SPTM2-ISTANet+ at different CRs for indoor and outdoor scenarios using the 
Cost2100 model. These two datasets
are commonly used in the CSI feedback works \cite{dl_multires, guo2019convolutional,tang2021dilated, ref:SRNet}.
The number of iteration blocks is set to 9. As shown in Fig. \ref{fig:nmse_init}(a)(b), our proposed SPTM2-ISTANet+ can achieve superior performance at all tested CRs. 
In particular, for CR of ${1}/{4}$ in the 
challenging outdoor scenario, SPTM2-ISTANet+
provides approximately $7$ dB improvement in CSI reconstruction accuracy compared to 
traditional DL-based schemes STNet, DCRNet, TransNet and CsiNet+. 
On the other hand, we notice that TiLISTA-Joint can perform better than traditional DL-based schemes for larger
CR (such as ${1}/{4}$).
The performance of TiLISTA-Joint is worse than that of traditional DL-based schemes at the higher compression (i.e. smaller CR), e.g., ${1}/{32}$, while SPTM2-ISTANet+ does not, demonstrating the effectiveness of our spherical processing and the residual design.}

\begin{figure}[!htb] 
\hspace*{-3mm}
	\subfigure[Cost2100 Indoor] {\label{fig:iter_indoor} 
	\includegraphics[width=0.25\textwidth]{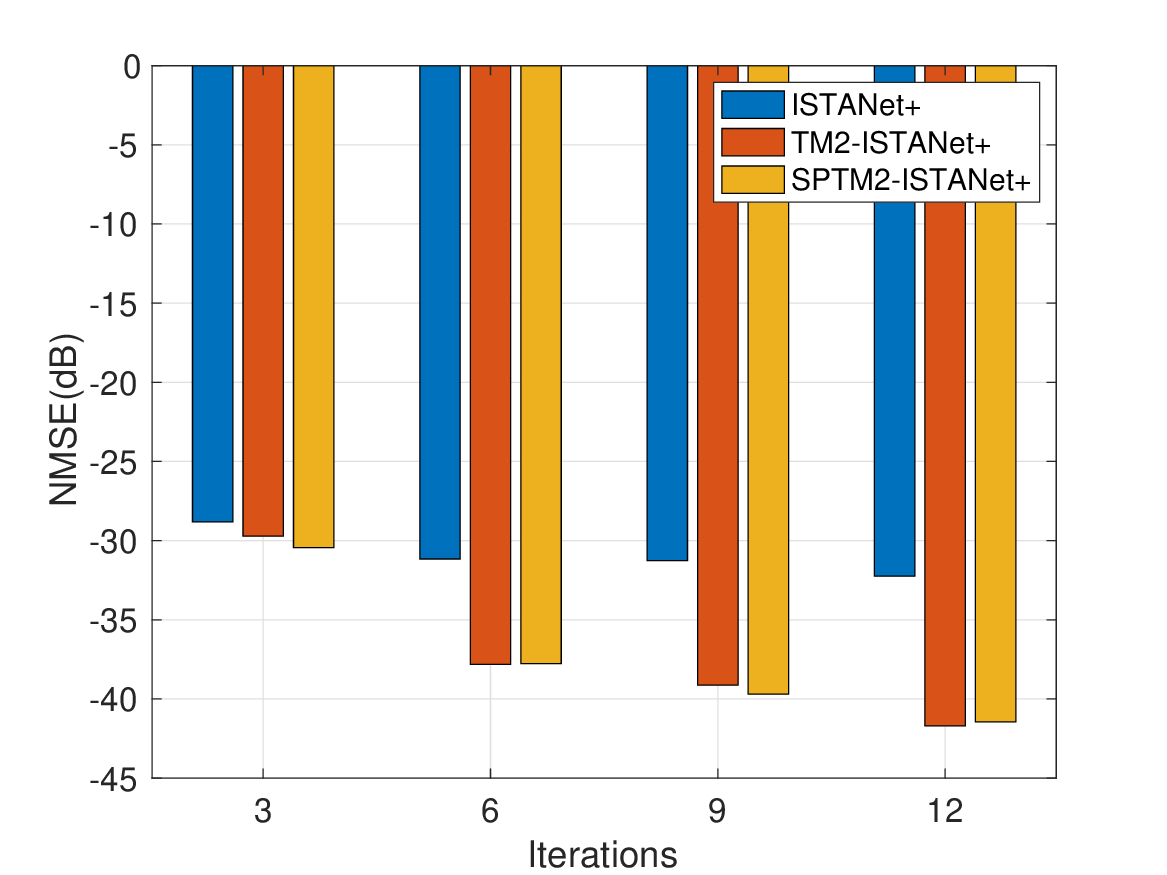}
	} \hspace*{-8mm}
	\subfigure[Cost2100 Outdoor] { \label{fig:iter_outdoor} 
	\includegraphics[width=0.25\textwidth]{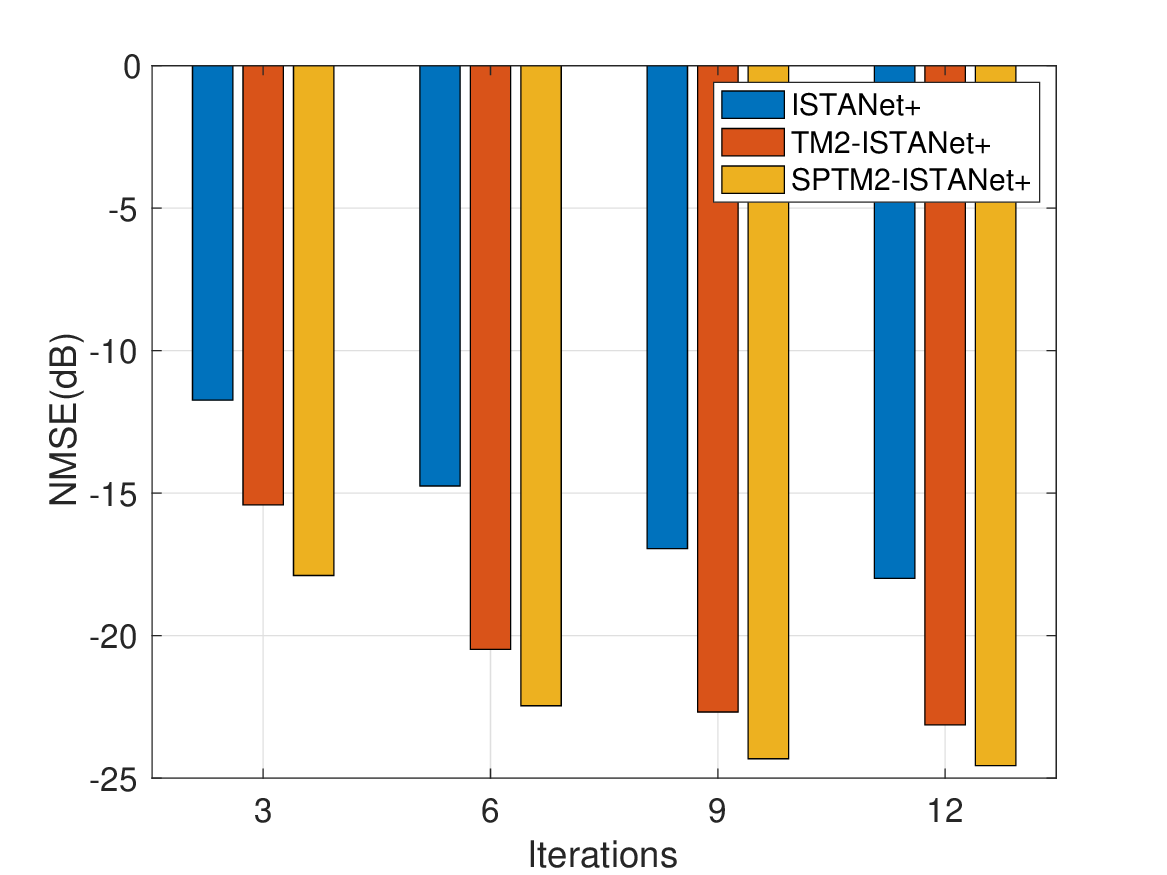} 
	} 
	\hspace*{-5mm}
        \caption{NMSE at different iteration blocks for CR = ${1}/{4}$.} 
	\label{fig:nmse_iter_org}  \vspace{-2mm}
\end{figure}

Fig. \ref{fig:nmse_iter_org} shows the impact of the number of iteration blocks $N_I$ on the CSI feedback performance of ISTANet+, TM2-ISTANet+ and SPTM2-ISTANet+ when CR= ${1}/{4}$ in indoor and outdoor scenarios against Cost2100 channel model.  TM2-ISTANet+ is the SPTM2-ISTANet+ without spherical processing to showcase the advantage of different components in SPTM2-ISTANet+.
We can observe that SPTM2-ISTANet+ delivers the best performance at various iterations. In fact, with
only 3 iterations, 
it achieves similar performance to ISTANet+ using
12-iterations. 
Evidently, the performance of SPTM2-ISTANet+ 
stabilizes after the number of iterative modules reaches 
9. Actually, 3 iteration modules would suffice for indoor channels at CR = ${1}/{4}$ since NMSE has already fallen below $-30$ dB, whereas 9 iterations 
would suffice for outdoor channels. The underlying reason is that channels and CSI variations in outdoor cases are relatively complex. Hence, more iterative modules are needed.

\subsection{Data Augmentation Comparison}

Since DL-based CSI feedback works have already achieved satisfactory performance for the indoor scenario \cite{dl_multires, guo2019convolutional,tang2021dilated, ref:SRNet}, we focus on the outdoor scenario where practical CSI measurement is even harder and CSI recovery  is less accurate.
For data augmentation, samples from the limited measurements will be randomly selected from the training set in Cost2100 Outdoor. We set shift ranges in the angular domain and delay domain to $-15$ to $15$ and $-3$ to $3$, respectively. We set the training dataset size after augmentation to the same overall training set size by repetition or phase randomization if the initial augmentation is not large enough (depending on the corresponding augmentation method). 

\begin{figure} \centering 
\includegraphics[width=0.7\columnwidth]{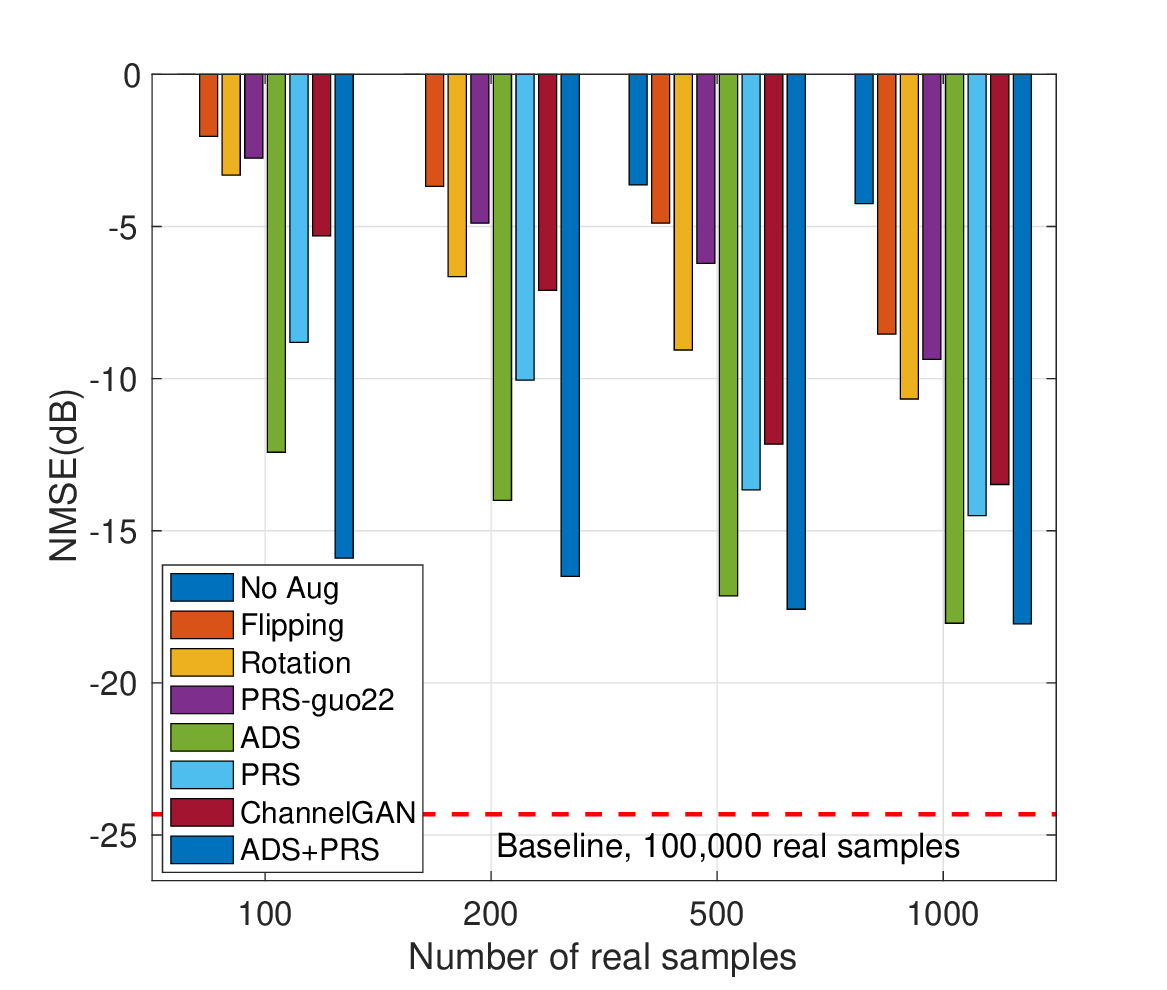} 
\caption{Augmentation comparison at CR = ${1}/{4}$.} 
\label{fig:figure_aug} \vspace*{-2mm}
\end{figure}

\textcolor{black}{Fig. \ref{fig:figure_aug} compares the performance of SPTM2-ISTANet+ using different augmentation strategies including the costly blackbox ``ChannelGAN'' \cite{ref:GAN_xiao}, no augmentation (``No Aug'') which uses repetition to enlarge dataset size, flipping, rotation, shift in angular-delay domain (``ADS'') where repetition is used to enlarge dataset size
only if necessary, single-value random phase shift (``PRS-guo22'') \cite{guo2022} which makes all elements in the CSI matrix share the same phase shift, random phase shift (``PRS''), and ADS together with PRS (ADS+PRS). We set the CR= ${1}/{4}$. 
We limit the number of CSI measurements before augmentation to $100$, $200$, $500$, and $1000$, respectively. As shown in Fig.~\ref{fig:figure_aug}, our proposed ADS and ADS+PRS significantly outperform ChannelGAN and other augmentation methods in each case. In fact,
PRS alone already outperforms ChannelGAN. Besides, our proposed PRS can bring about $5$ dB gain compared with PRS-guo22 by considering the variation of phase shift in different paths.
Notably, ADS+PRS can achieve NMSE of $-15.8$ dB using only $100$ CSI measurement samples whereas ChannelGAN reaches $-5.3$ dB. Furthermore, both ADS and PRS
improve CSI recovery accuracy.
Overall, the proposed low-cost training enhancement by using ADS always achieves higher gains than PRS owing to the better utilization of geographical correlation
based on domain knowledge.}

\subsection{Recovery Comparison against New Scenarios}

To evaluate recovery performance when UE encounters a new scenario, we select the model pretrained from Cost2100 Indoor, Cost2100 Outdoor, Measured Indoor and Quadriga 3GPP UMA, respectively. For each model,
we use the other three datasets unseen by
the pretrained model as new scenarios. 

\begin{figure} \centering 
\includegraphics[width=1.05\columnwidth]{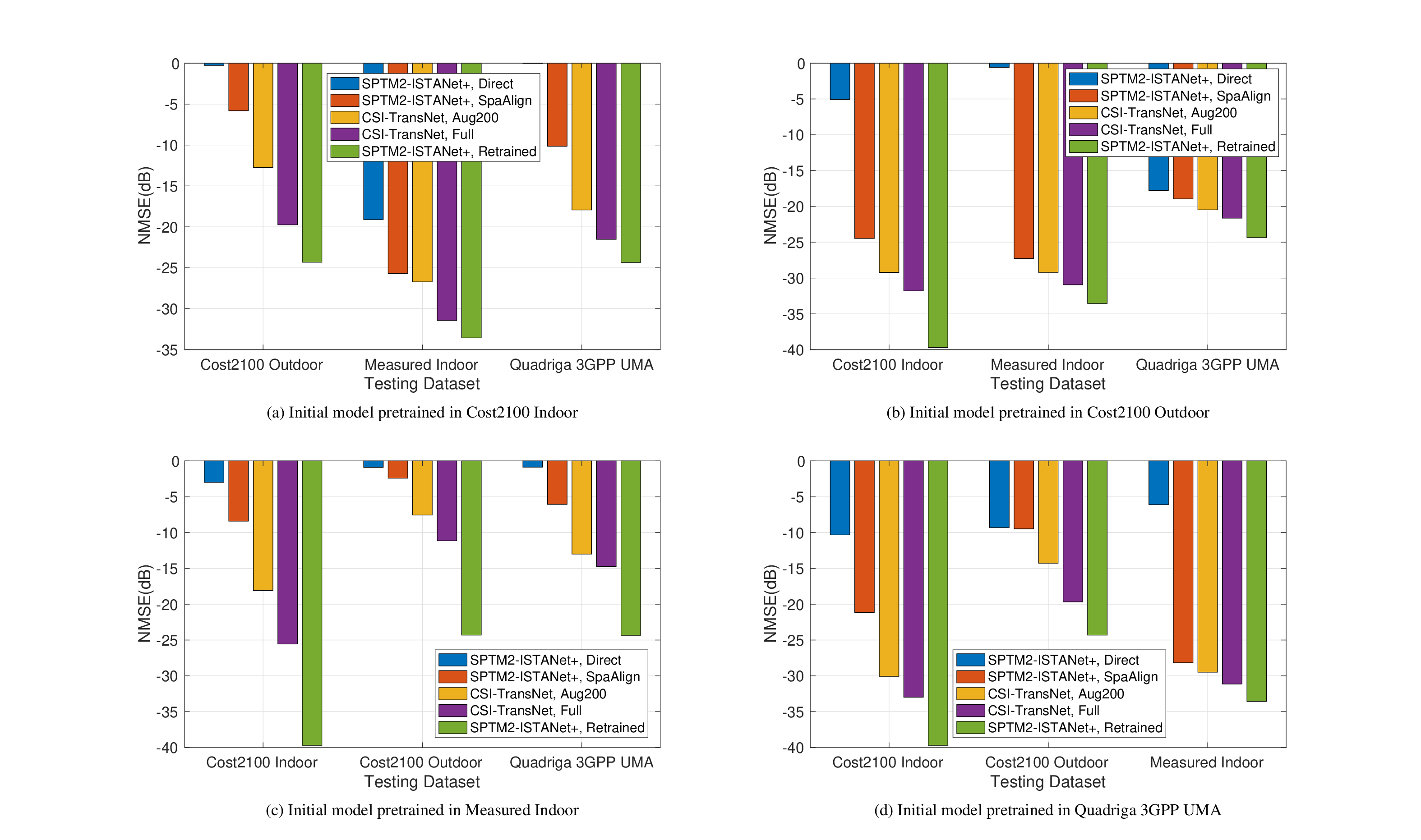} 
\caption{NMSE comparison for the model pretrained in one scenario to the new scenarios when CR = ${1}/{4}$.} 
\label{fig:figure_trans_all_cr4} \vspace*{-4mm}
\end{figure}

\begin{table*}[h!]
\centering
\color{black}
\caption{\textcolor{black}{Cosine similarity comparison for the model pretrained in one scenario to the new scenarios when CR = ${1}/{4}$.} }
\label{tab:combined_scenarios_cos}
\begin{tabular}{|c|c|c|c|c|c|c|}
\hline
\textbf{Pretrained Scenario} & \textbf{New Scenario} & \textbf{Direct} & \textbf{SpaAlign} & \textbf{Aug200} & \textbf{Full} & \textbf{Retrained} \\ \hline
\multirow{3}{*}{\textbf{Cost2100 Indoor}} & \textbf{Cost2100 Outdoor} & 0.33 & 0.84 & 0.95 & 0.97 & 0.98 \\ \cline{2-7}
 & \textbf{Measured Indoor} & 0.99 & 1.00 & 1.00 & 1.00 & 1.00 \\ \cline{2-7}
 & \textbf{Quadriga 3GPP UMA} & 0.28 & 0.95 & 0.99 & 0.99 & 0.99 \\ \hline
\multirow{3}{*}{\textbf{Cost2100 Outdoor}} & \textbf{Cost2100 Indoor} & 0.87 & 0.99 & 1.00 & 1.00 & 1.00 \\ \cline{2-7}
 & \textbf{Measured Indoor} & 0.51 & 1.00 & 1.00 & 1.00 & 1.00 \\ \cline{2-7}
 & \textbf{Quadriga 3GPP UMA} & 0.99 & 0.99 & 0.99 & 0.99 & 0.99 \\ \hline
\multirow{3}{*}{\textbf{Measured Indoor}} & \textbf{Cost2100 Indoor} & 0.80 & 0.94 & 0.99 & 1.00 & 1.00 \\ \cline{2-7}
 & \textbf{Cost2100 Outdoor} & 0.47 & 0.65 & 0.89 & 0.94 & 0.98 \\ \cline{2-7}
 & \textbf{Quadriga 3GPP UMA} & 0.46 & 0.86 & 0.97 & 0.98 & 0.99 \\ \hline
\multirow{3}{*}{\textbf{Quadriga 3GPP UMA}} & \textbf{Cost2100 Indoor} & 0.96 & 0.99 & 1.00 & 1.00 & 1.00 \\ \cline{2-7}
 & \textbf{Cost2100 Outdoor} & 0.92 & 0.92 & 0.96 & 0.97 & 0.98 \\ \cline{2-7}
 & \textbf{Measured Indoor} & 0.89 & 1.00 & 1.00 & 1.00 & 1.00 \\ \hline
\end{tabular}
\end{table*}

\begin{figure*}[!htb] \centering 
	\subfigure[Cost2100 Indoor] {\label{fig:trans_outdoor2indoor} 
	\includegraphics[width=0.3\textwidth]{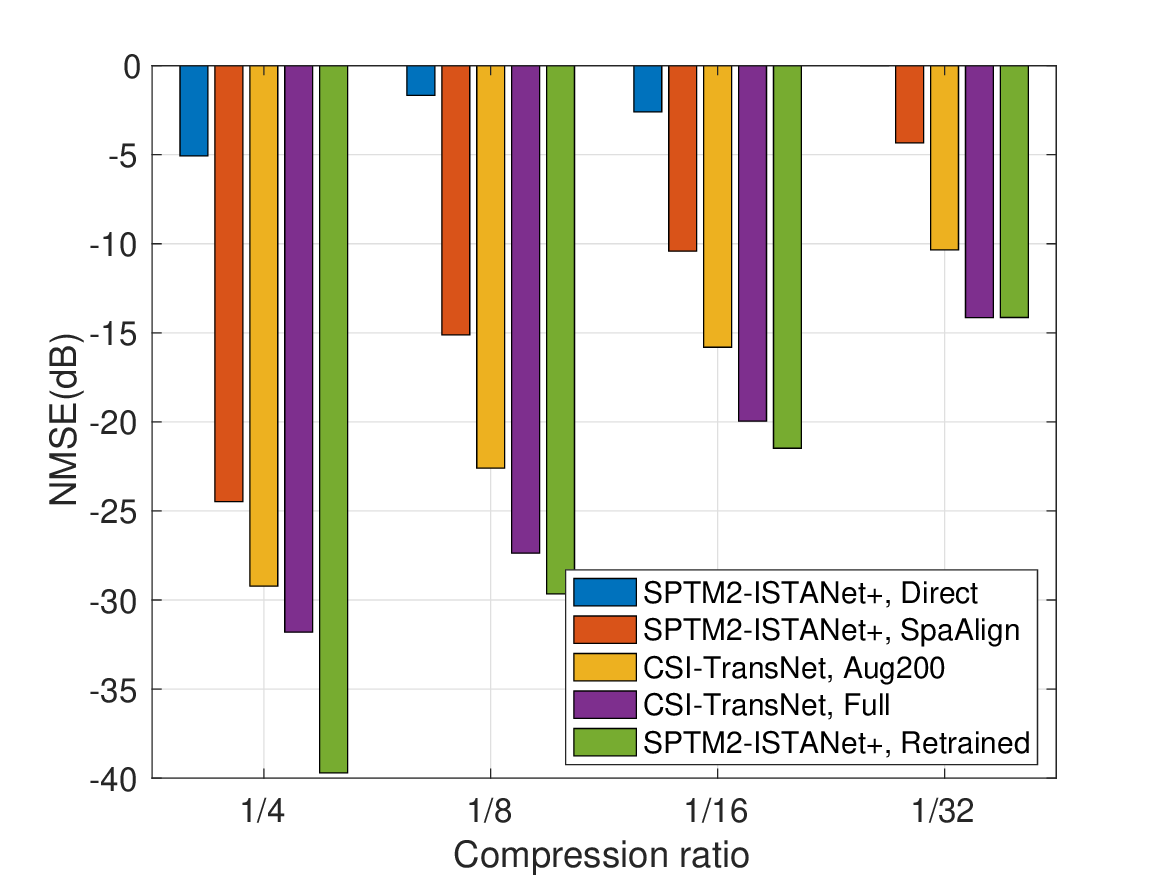}
	} 
	\subfigure[Measured Indoor] { \label{fig:trans_outdoor2ula} 
	\includegraphics[width=0.3\textwidth]{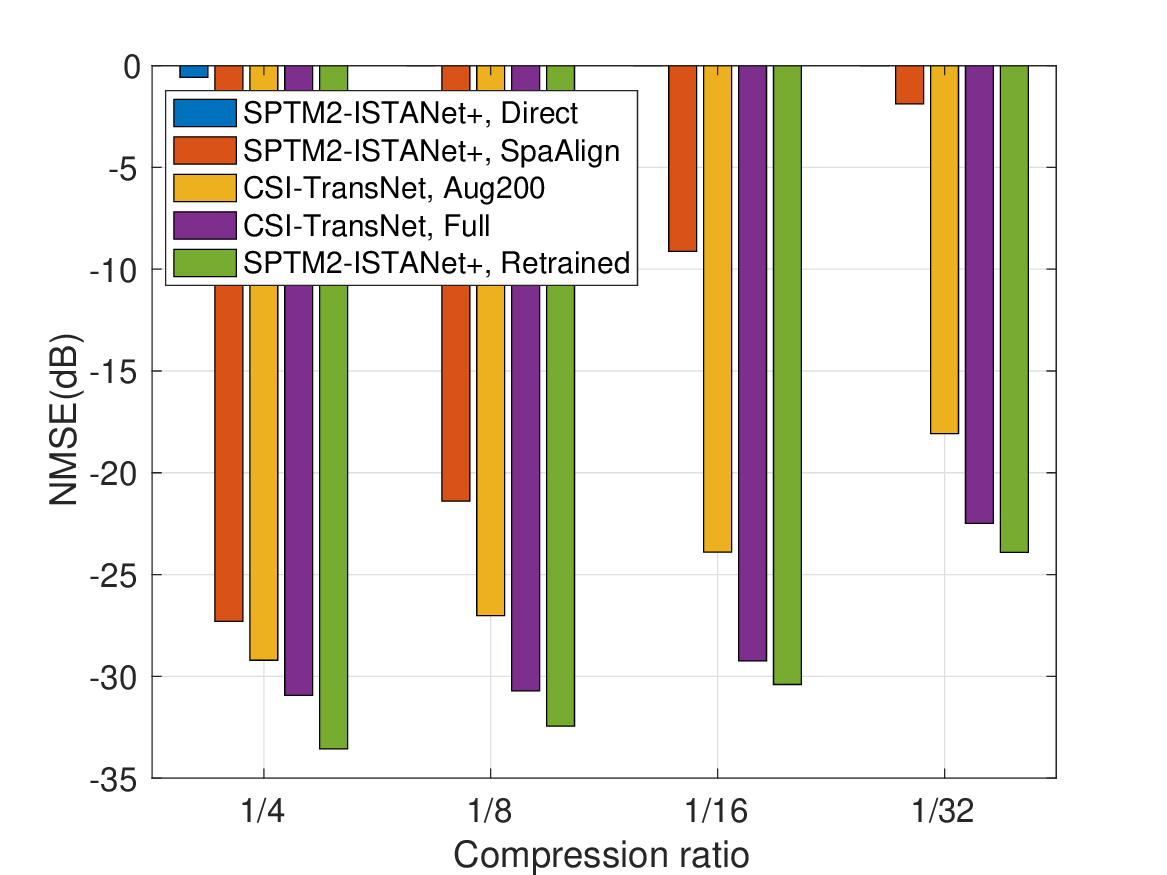} 
	} 
        \subfigure[Quadriga 3GPP UMA] { \label{fig:trans_outdoor2quadriga} 
	\includegraphics[width=0.3\textwidth]{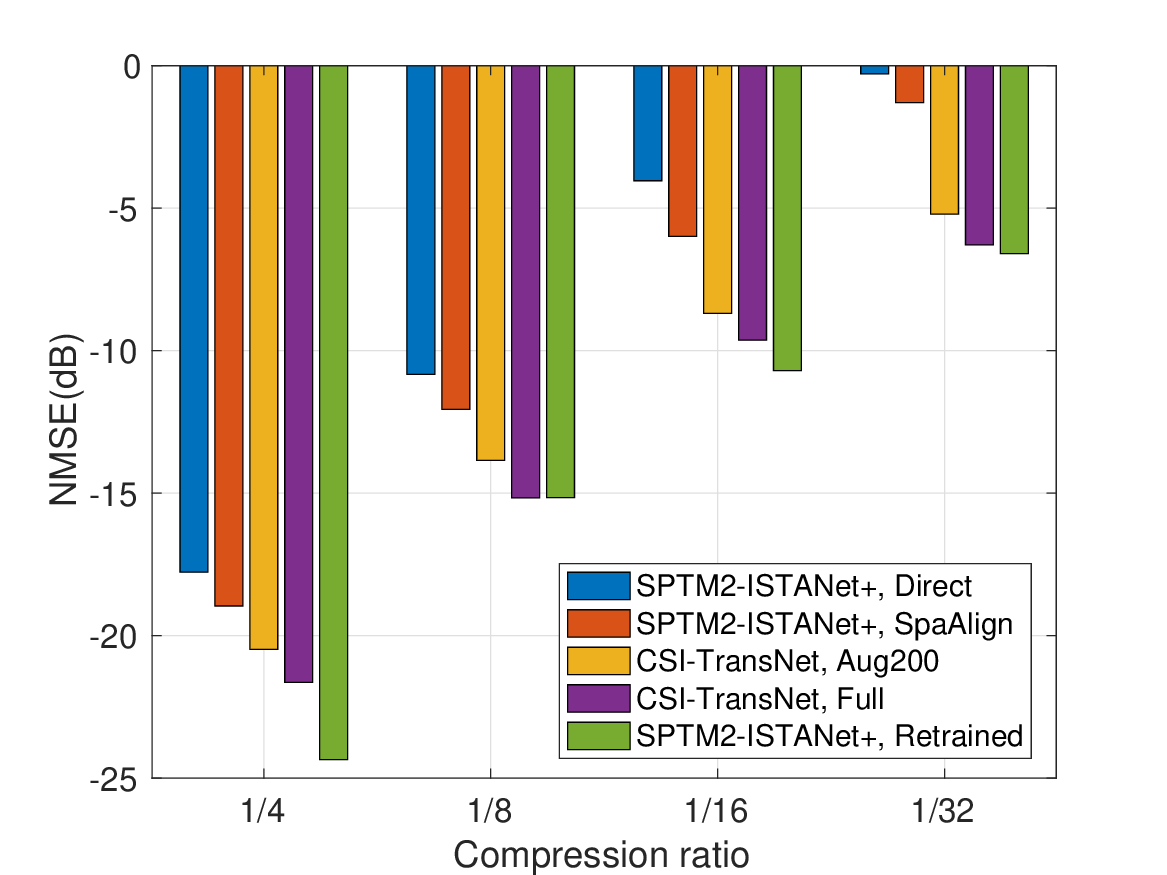} 
	} 
	\vspace*{-3mm}
        \caption{NMSE comparison in different CRs for the model pretrained in the Cost2100 Outdoor dataset to the new scenarios.} 
	\label{fig:nmse_trans_outdoor} 
\end{figure*}

\begin{table*}[h!]
\centering
\color{black}
\caption{ \textcolor{black}{Cosine similarity comparison in different CRs for the model pretrained in the Cost2100 Outdoor dataset to the new scenarios.}}
\label{tab:combined_scenarios_cos_outdoor}
\begin{tabular}{|c|c|c|c|c|c|c|}
\hline
\textbf{Scenario} & \textbf{ } & \textbf{Direct} & \textbf{SpaAlign} & \textbf{Aug200} & \textbf{Full} & \textbf{Retrained} \\ \hline
\multirow{4}{*}{\textbf{Cost2100 Indoor}} & \textbf{CR=$\frac {1}{4}$}  & 0.87 & 0.99 & 1.00 & 1.00 & 1.00 \\ \cline{2-7}
 & \textbf{CR=$\frac {1}{8}$}  & 0.69 & 0.98 & 0.99 & 1.00 & 1.00 \\ \cline{2-7}
 & \textbf{CR=$\frac {1}{16}$} & 0.69 & 0.96 & 0.98 & 0.99 & 0.99 \\ \cline{2-7}
 & \textbf{CR=$\frac {1}{32}$} & 0.19 & 0.81 & 0.96 & 0.98 & 0.98 \\ \hline
\multirow{4}{*}{\textbf{Measured Indoor}} & \textbf{CR=$\frac {1}{4}$}  & 0.51 & 1.00 & 1.00 & 1.00 & 1.00 \\ \cline{2-7}
 & \textbf{CR=$\frac {1}{8}$}  & 0.24 & 1.00 & 1.00 & 1.00 & 1.00 \\ \cline{2-7}
 & \textbf{CR=$\frac {1}{16}$} & 0.18 & 0.94 & 1.00 & 1.00 & 1.00 \\ \cline{2-7}
 & \textbf{CR=$\frac {1}{32}$} & 0.17 & 0.62 & 0.99 & 1.00 & 1.00 \\ \hline
\multirow{4}{*}{\textbf{Quadriga 3GPP UMA}} & \textbf{CR=$\frac {1}{4}$}  & 0.99 & 0.99 & 0.99 & 0.99 & 0.99 \\ \cline{2-7}
 & \textbf{CR=$\frac {1}{8}$}  & 0.95 & 0.96 & 0.97 & 0.98 & 0.98 \\ \cline{2-7}
 & \textbf{CR=$\frac {1}{16}$} & 0.78 & 0.86 & 0.93 & 0.94 & 0.94 \\ \cline{2-7}
 & \textbf{CR=$\frac {1}{32}$} & 0.32 & 0.53 & 0.83 & 0.87 & 0.85 \\ \hline
\end{tabular}
\end{table*}

\textcolor{black}{Fig. \ref{fig:figure_trans_all_cr4} shows CSI feedback performance for the model pretrained in one scenario to the new scenarios at CR = ${1}/{4}$.  ``SPTM2-ISTANet+, Direct" denotes the performance of the pretrained SPTM2-ISTANet+ in the new scenario without preprocessing,  ``SPTM2-ISTANet+, SpaAlign" shows the performance of the pretrained SPTM2-ISTANet+ with the help of the sparsity aligning function $f_\text{sa}(\cdot)$, ``CSI-TransNet, Aug200" 
indicates the performance of CSI-TransNet with the help of the proposed ADS+PRS augmentation 
starting with 200 samples from the new scenario.
As baselines, ``CSI-TransNet, Full" corresponds to CSI-TransNet 
trained with the full training sets in the new scenario, and ``SPTM2-ISTANet+, Retrained" corresponds to the performance of SPTM2-ISTANet+ that 
are retrained from scratch in the new scenario.}

As shown in Fig. \ref{fig:figure_trans_all_cr4}, ``SPTM2-ISTANet+, Direct" generally suffers poor CSI recovery accuracy, which is consistent with
common knowledge
that DL-based CSI feedback has a limited ability to handle more general CSIs. With the help of sparsity aligning, the models pretrained on Cost2100 Outdoor and Quadriga 3GPP UMA demonstrate clear accuracy improvement in the new scenarios. 
Actually, compared with Cost2100 Indoor and Measured Indoor, CSI in Cost2100 Outdoor and Quadriga 3GPP UMA exhibit more diverse features in terms of
multipath delay and angle of arrival/departure distributions. This allows models pretrained in more diverse environments to tackle more general
channel scenarios. 

With limited measurement data,
``CSI-TransNet, Aug200" illustrates obvious performance improvement for each case,
especially for the anchor model pretrained on Cost2100 Indoor and Measured Indoor.
This observation confirms the effectiveness of our proposed plug-in translation module and model-driven channel data augmentation. 
``CSI-TransNet, Full" provides additional gains over ``CSI-TransNet, Aug200", which means that a more accurate CSI-to-CSI translation can benefit more from the pretrained model. ``SPTM2-ISTANet+, Retrained" is selected as the performance bound for the CSI feedback accuracy.
We observe that the models pretrained on Cost2100 Outdoor exhibit smaller gaps to the retraining bound. Consequently, we suggest using the model pretrained in a more diverse (complex) environment
(such as outdoor) to serve as the anchor network in CSI-TransNet.

 \textcolor{black}{
Table \ref{tab:combined_scenarios_cos} presents the cosine similarity of models pretrained in one scenario when applied to new scenarios at CR = ${1}/{4}$. For brevity, the terms ``SPTM2-ISTANet+, Direct", ``SPTM2-ISTANet+, SpaAlign", ``CSI-TransNet, Aug200", ``CSI-TransNet, Full", and ``SPTM2-ISTANet+, Retrained" are abbreviated as ``Direct", ``SpaAlign", ``Aug200", ``Full", and ``Retrained", respectively.
As depicted in Table \ref{tab:combined_scenarios_cos}, and consistent with the NMSE results, our proposed plug-in translation module and model-driven channel data augmentation yield effective precoding gains compared to the direct reuse of the pretrained model in new scenarios, and anchor models pretrained on Cost2100 Outdoor overall register the best cosine similarity performance. Furthermore, the pairings of Cost2100 Indoor with Measured Indoor, and Cost2100 Outdoor with Quadriga 3GPP UMA demonstrate enhanced performance when reused, highlighting the potential to classify wireless environments based on similar features for model reuse to minimize model update overheads.
}

\textcolor{black}{We now consider the CSI recovery for different CR levels.
Fig. \ref{fig:nmse_trans_outdoor} and Table \ref{tab:combined_scenarios_cos_outdoor} present the NMSE and cosine similarity for the model pretrained using the Cost2100 Outdoor dataset across different scenarios, respectively. A notable observation is that ``CSI-TransNet, Full" consistently exhibits comparable CSI recovery accuracy and cosine similarity to ``SPTM2-ISTANet+, Retrained" in different CRs. This suggests that our proposed plug-in CSI-to-CSI translation design is able
to reuse the pretrained model and weights effectively.
Furthermore, the ``CSI-TransNet, Aug200" demonstrates robust performance, achieving an NMSE below $-20$ dB at CR = ${1}/{4}$ and below $-8.6$ dB at CR = ${1}/{16}$ in various scenarios. Correspondingly, it attains a cosine similarity exceeding $0.99$ at CR = ${1}/{4}$ and above $0.93$ at CR = ${1}/{16}$. These results highlight the capability of ``CSI-TransNet, Aug200" as a versatile anchor network, and suggests that the anchor network pretrained on a diverse environment benefits scenario-adaptive design.
}

Interestingly, note that carrier frequencies of Cost2100 Indoor, Measured Indoor and Quadriga 3GPP UMA are different from those of Cost2100 Outdoor. In fact, the subcarrier spacing of Measured Indoor is also different from Cost2100 Outdoor. The performance robustness of 
CSI-TransNet irrespective of the difference
in terms of carrier
frequencies and subcarrier spacing shows that our proposed CSI-TransNet architecture is less rigid and can reuse
pretrained networks across different scenarios.
Using very limited new measurements from a new
environment, CSI-TransNet reduces the cost of practical deployment in diverse and complex wireless environments.

\addtolength{\topmargin}{0.03in}
\subsection{Robustness Evaluation}

\begin{figure}[!htb] 
\hspace*{-3mm}
	\subfigure[CR = ${1}/{4}$] {\label{fig:quan_cr4} 
	\includegraphics[width=0.25\textwidth]{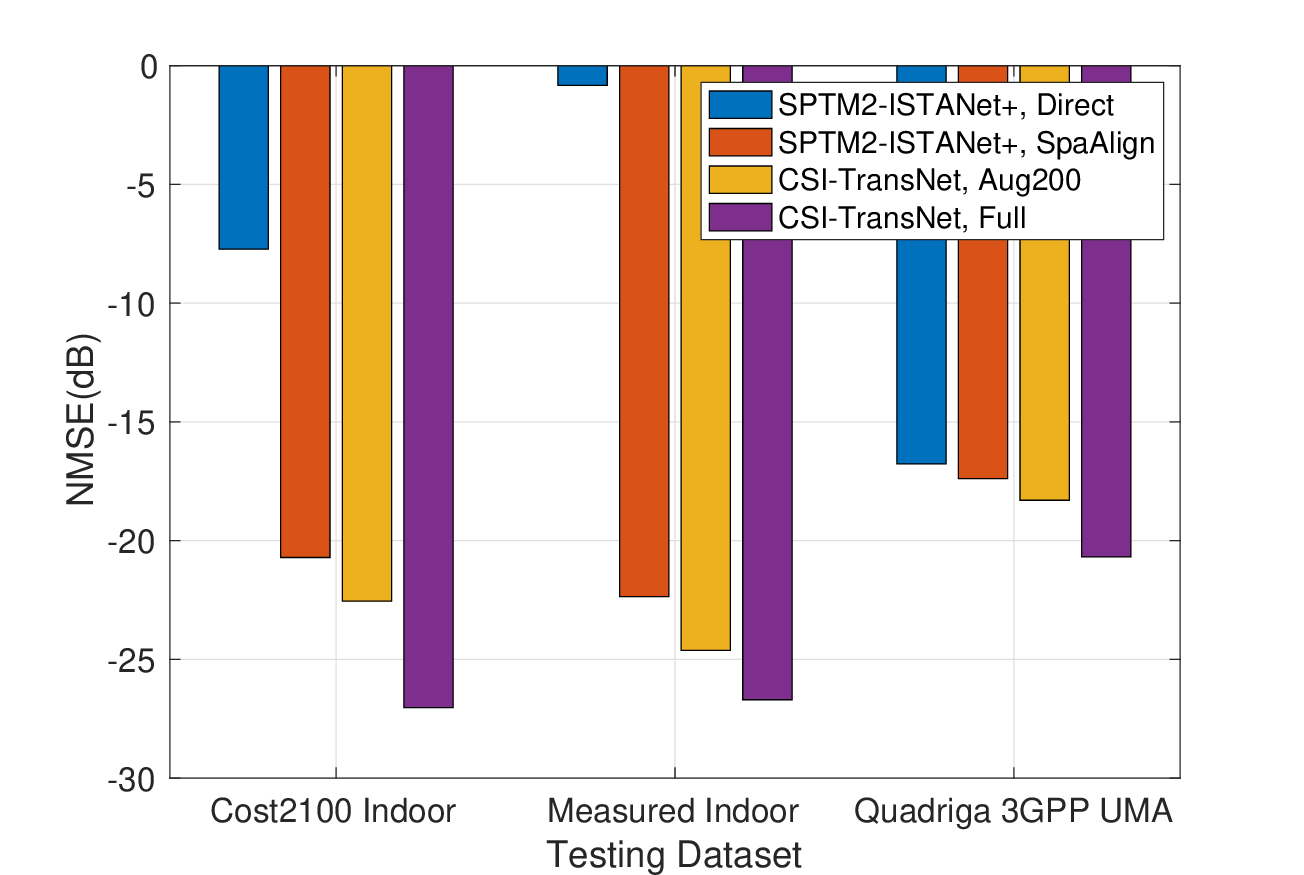}
	} \hspace*{-8mm}
	\subfigure[CR = ${1}/{16}$] { \label{fig:quan_cr16} 
	\includegraphics[width=0.25\textwidth]{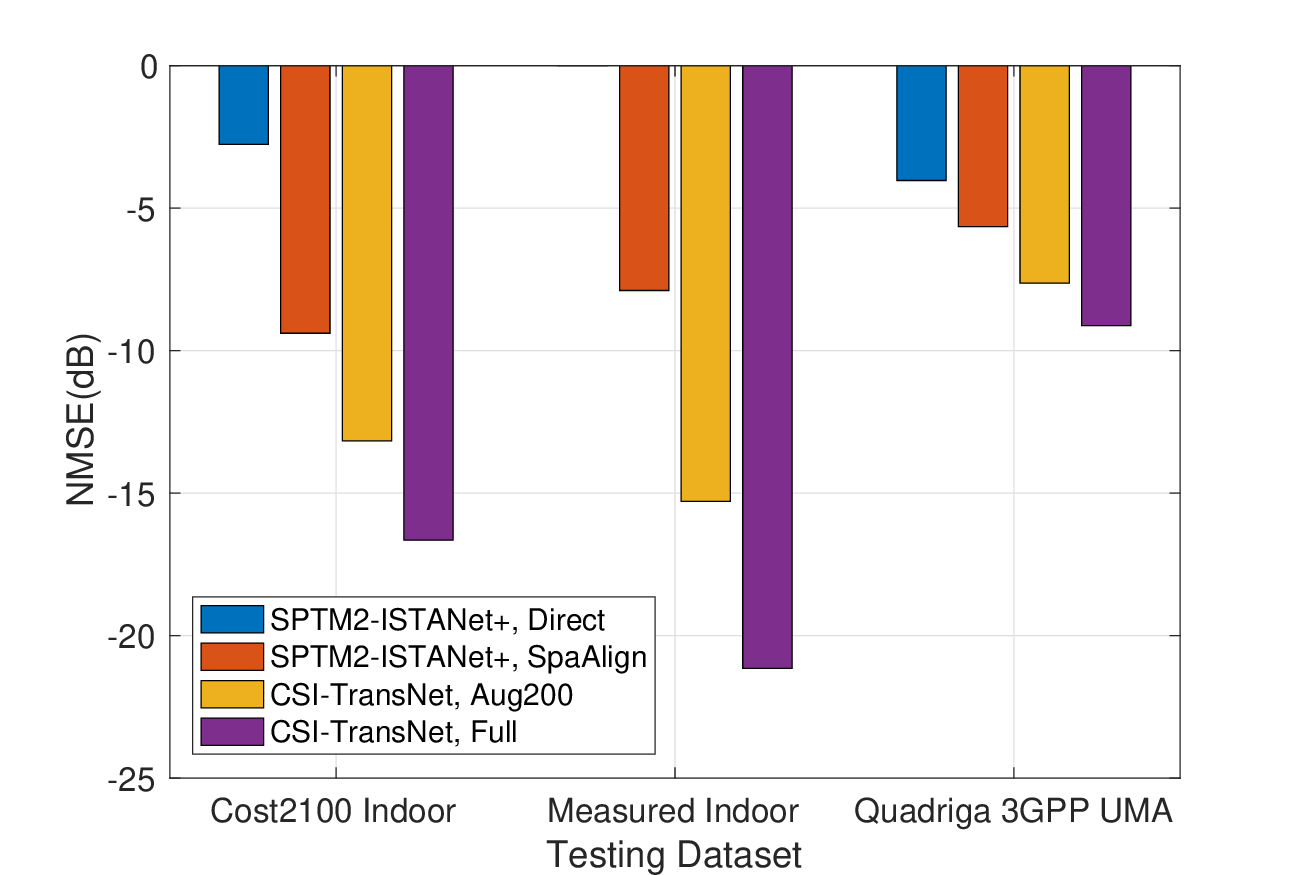} 
	} 
	\hspace*{-3mm}
        \caption{NMSE at 4-bit quantization for the model pretrained in the Cost2100 Outdoor dataset to the new scenarios.} 
	\label{fig:nmse_iter}  \vspace{-4mm}
\end{figure}

\textcolor{black}{In this subsection, we further evaluate the robustness of CSI-TransNet considering the influence of codewords quantization and channel noise.}

\textcolor{black}{The importance of CSI feedback being conveyed as a bitstream necessitates the consideration of quantization within the process. While there exist DL-based CSI feedback works that incorporate quantization \cite{lu2020bit, liu2020quan}, we specifically evaluate its impact within our proposed CSI-TransNet to discern the effects of feedback quantization. We employ the $\mu$-law quantization technique presented in \cite{liu2020quan}, applying it to the dimension-compressed codewords of our backbone network, SPTM2-ISTANet+. Subsequent fine-tuning is performed to adapt the network to the quantization.  Post 4-bit quantization, SPTM2-ISTANet+ reports NMSE values of $-20.2$ dB for CR = ${1}/{4}$ and $-7.3$ dB for CR = ${1}/{16}$ in the Cost2100 Outdoor setting. After that, we test the performance of 4-bit quantization for the backbone model pretrained in the Cost2100 Outdoor scenario to the new scenarios. }

\textcolor{black}{Based on the model pretrained at 4-bit quantization for the Cost2100 Outdoor scenario, Fig. \ref{fig:quan_cr4} and Fig. \ref{fig:quan_cr16} present the NMSE performance for CSI-TransNet across different scenarios, respectively. 
Notably, ``CSI-TransNet, Full" consistently demonstrates marked improvements in CSI recovery accuracy for varying CRs compared to a direct reuse of the pretrained network. This reinforces the efficacy of our proposed plug-in CSI-to-CSI translation design in leveraging pretrained models and their weights. Moreover, ``CSI-TransNet, Aug200" records an NMSE below $-18.0$ dB and $-7.6$ dB at CR = ${1}/{4}$ and CR = ${1}/{16}$ across scenarios, underlining the resilience and robustness of CSI-TransNet amidst quantization effects.}

  \begin{figure}[!htb] \centering 
	\subfigure[ Generalizability of SPTM2-ISTANet+ pretrained in noisy Cost2100 outdoor] {\label{fig:denoise_a} 
	\includegraphics[width=0.34\textwidth]{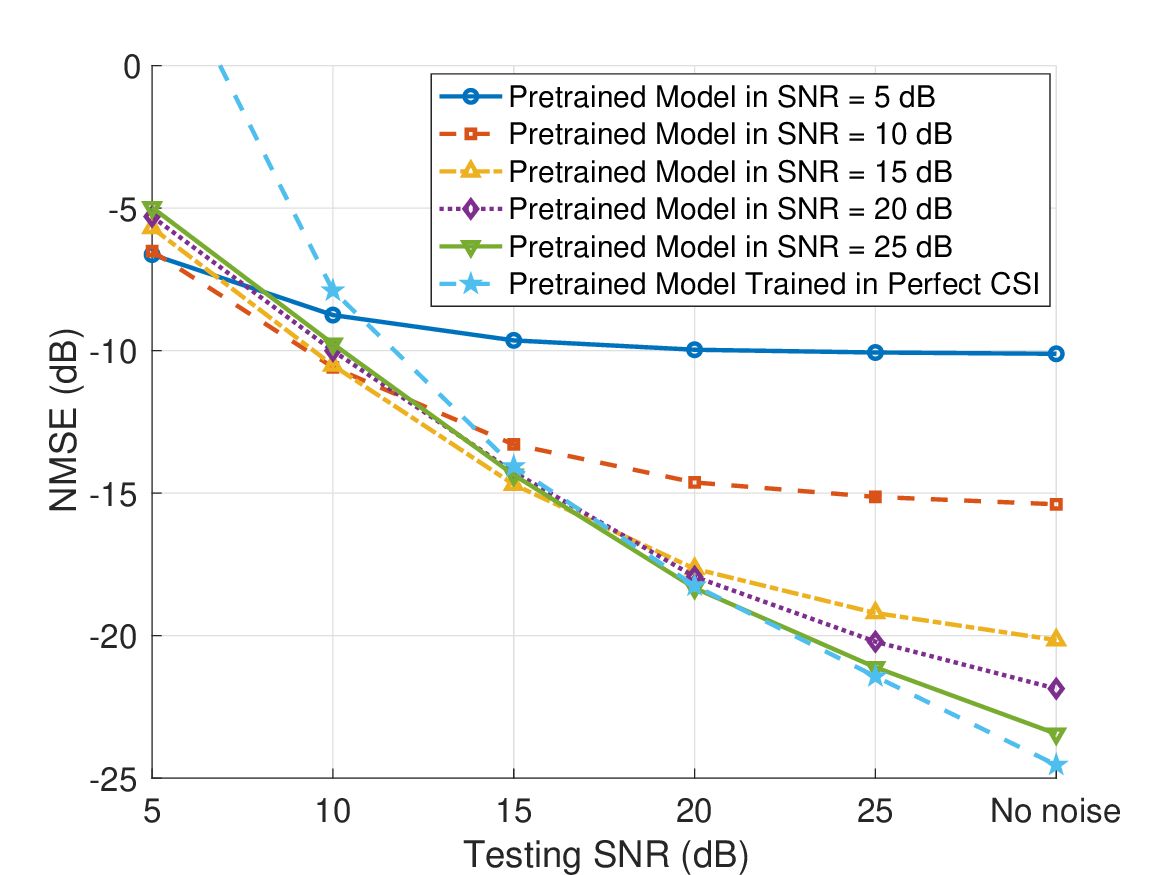}
	}
	\subfigure[Performance evaluation of CSI-TransNet using the backbone model pretrained in Cost2100 outdoor SNR = 25dB ] { \label{fig:denoise_b} 
	\includegraphics[width=0.34\textwidth]{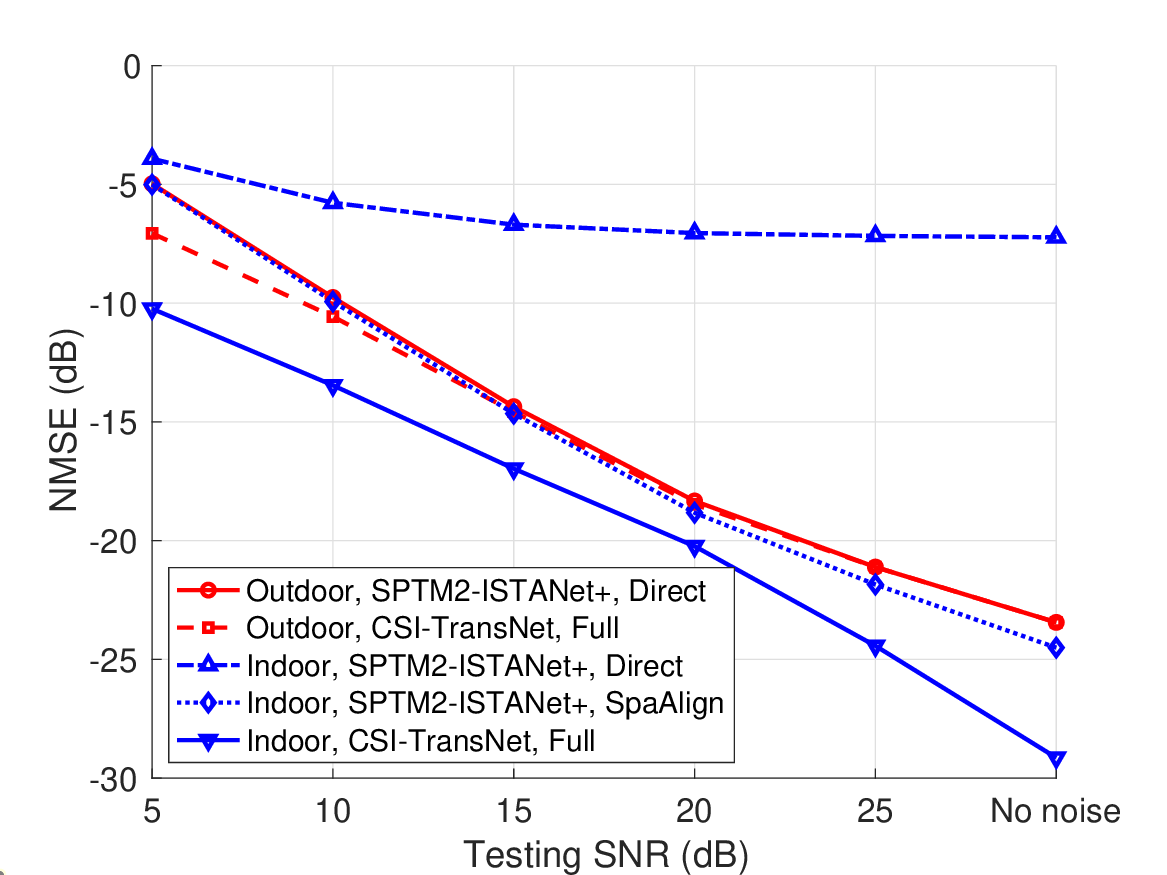} 
	} 
	\vspace*{-3mm}
        \caption{Evaluation of SPTM2-ISTANet+ and CSI-TransNet over channel SNR variations ranges.} 
	\label{fig:nmse_snr} \vspace*{-3mm}
\end{figure}

\textcolor{black}{To further test the robustness of the backbone network SPTM2-ISTANet+ and CSI-Transnet framework, we examine the reconstruction performance in the presence of channel estimation error owing to the influence of channel noise. Fig. \ref{fig:denoise_a} studies the generalizability of the pretrained SPTM2-ISTANet+ when there is a mismatch between the training and testing signal-to-noise ratios (SNRs). Taking the Cost2100 outdoor scenario as an example, we compare the massive MIMO CSI recovery performance at different SNR values when CR $=1/4$. As shown in Fig. \ref{fig:denoise_a}, the backbone network suffers from obvious performance degradation when the testing SNR falls below the SNR value used for training, especially for the model pretrained in the ideal channel.
 Besides, for a model pretrained at a lower SNR, performance plateaus when the test SNR exceeds the training value. Notably, the model pretrained in SNR $ = 25$ dB in the Cost2100 outdoor scenario has a good anti-noise characteristic and high recovery accuracy in the high SNR cases, which means it can be a backbone model over a wider SNR range and new environment. }
 
\textcolor{black}{Fig. \ref{fig:denoise_b} evaluates the robustness of the CSI-Transnet framework when selecting the model pretrained in the Cost2100 outdoor scenario at SNR $= 25$ dB and CR $= 1/4$ as the backbone model in the new scenario and SNR. Actually, the translation model can not only perform the cross-environment CSI-to-CSI translation but also denoising. As shown in Fig. \ref{fig:denoise_b}, taking the model pretrained in SNR $ = 25$ dB in the Cost2100 outdoor scenario as the backbone model, both sparsity aligning and CSI-to-CSI translation in the CSI-TransNet framework can bring obvious gains in the Cost2100 indoor noisy scenario, which means CSI-TransNet can be robust considering the SNR-adaptive requirement. We can also note that the recovery accuracy of CSI-TransNet can achieve $-10$ dB even when the SNR $ = 5$ dB in the Cost2100 indoor scenario.
}

\subsection{Complexity Comparison}
\vspace*{-1mm}

\begin{table*}[]
\renewcommand{\arraystretch}{1.5}
\color{black}
\centering
\caption{\textcolor{black}{Model size (Parameters) of encoder networks in UE. M: million, K: thousand.}}
\label{tab:params_complex}
\begin{tabular}{|c|c|c|c|c|c|}
\hline
& \textbf{DCRNet} & \textbf{STNet} & \textbf{TransNet} & \textbf{SPTM2-ISTANet+} & \textbf{TransModule} \\ \hline
\textbf{CR=$\frac {1}{4}$} & 1.0 M & 1.1 M & 1.3 M & 1.0 M & 1.8 K \\ \hline
\textbf{CR=$\frac {1}{8}$} & 0.5 M & 0.5 M & 0.8 M & 0.5 M & 1.8 K \\ \hline
\textbf{CR=$\frac {1}{16}$} & 0.3 M & 0.3 M & 0.5 M & 0.3 M & 1.8 K \\ \hline
\textbf{CR=$\frac {1}{32}$} & 0.1 M & 0.1 M & 0.4 M & 0.1 M & 1.8 K \\ \hline
\end{tabular}
\end{table*}

\begin{table*}[]
\renewcommand{\arraystretch}{1.5}
\color{black}
\centering
\caption{\textcolor{black}{Computational complexity (FLOPs) of encoder networks in UE. M: million.}}
\label{tab:flops_complex}
\begin{tabular}{|c|c|c|c|c|c|}
\hline
& \textbf{DCRNet} & \textbf{STNet} & \textbf{TransNet} & \textbf{SPTM2-ISTANet+} & \textbf{TransModule} \\ \hline
\textbf{CR=$\frac {1}{4}$} & 2.6 M & 2.9 M & 17.9 M & 2.1 M & 3.7 M \\ \hline
\textbf{CR=$\frac {1}{8}$} & 1.6 M & 2.4 M & 17.3 M & 1.0 M & 3.7 M \\ \hline
\textbf{CR=$\frac {1}{16}$} & 1.0 M & 2.2 M & 17.1 M & 0.5 M & 3.7 M \\ \hline
\textbf{CR=$\frac {1}{32}$} & 0.8 M & 2.0 M & 16.9 M & 0.3 M & 3.7 M \\ \hline
\end{tabular}
\end{table*}

\begin{table*}[]
\renewcommand{\arraystretch}{1.5}
\centering
\color{black}
\caption{\textcolor{black}{Runtime of encoder networks (in \(10^{-5}\) second).}}
\label{tab:exec_time}
\begin{tabular}{|c|c|c|c|c|c|c|}
\hline
& \textbf{DCRNet} & \textbf{STNet} & \textbf{TransNet} & \textbf{SPTM2-ISTANet+} & \textbf{CSI-TransNet} & \textbf{TransModule} \\ \hline
\textbf{CR=$\frac {1}{4}$} & 0.5 & 1.2 & 1.3 & 0.1 & 0.8 & 0.8 \\ \hline
\textbf{CR=$\frac {1}{8}$} & 0.5 & 1.2 & 1.3 & 0.1 & 0.8 & 0.8 \\ \hline
\textbf{CR=$\frac {1}{16}$}& 0.5 & 1.2 & 1.3 & 0.1 & 0.8 & 0.8 \\ \hline
\textbf{CR=$\frac {1}{32}$}& 0.5 & 1.2 & 1.3 & 0.1 & 0.8 & 0.8 \\ \hline
\end{tabular}
\end{table*}

\textcolor{black}{We provide some discussions with respect to algorithm complexity. 
 Tables~\ref{tab:params_complex} and ~\ref{tab:flops_complex} outline the parameters and FLOPs comparison for the cost-sensitive UE encoder and translation module. We do not find it interesting to compare decoder networks because the computation power and energy of resource-rich gNBs 
are generally of less concern. 
The comparison shows that SPTM2-ISTANet+ can reduce UE computation by over $88\%$, $28\%$and $19\%$, respectively, in comparison with TransNet, STNet and DCRNet at CR = ${1}/{4}$.
Computation saving grows as CR decreases. The size
(number of parameters) of the above encoders
are at a similar level, except TransNet, which utilizes at least an additional 30$\%$ parameters. Remarkably, when juxtaposed with conventional encoder networks, our novel plug-in translation module drastically reduces the number of UE parameters needing updates in fresh scenarios, from a staggering one million to just 1.8 thousand.  Besides, TransNet's usage is exponentially higher, attributed to its transformer-based architecture, whereas SPTM2-ISTANet+ is more parsimonious. Despite the translation module's FLOPs being double that of SPTM2-ISTANet+ at CR = ${1}/{4}$, the convolutional layers' computational intricacy can be ameliorated by harnessing lightweight optimization techniques \cite{ref:lightweight1}, such as depth-wise convolution \cite{ref:lightweight2} and knowledge distillation \cite{ref:Knowledge_distillation}.}

\textcolor{black}{For a practical insight, Table~\ref{tab:exec_time} provides runtime comparisons among encoder networks. Tests conducted on the PyTorch platform using the Nvidia GeForce RTX 3080 GPU reveal that SPTM2-ISTANet+ boasts the shortest runtime, while TransNet lags behind with the longest. Additionally, the proposed CSI-TransNet framework, comprising TransModule and SPTM2-ISTANet+, exhibits superior runtime efficiency when set against TransNet and STNet, demonstrating the effectiveness of the proposed translation design.}

Table~\ref{tab:aug-comp-complex} compares parameters and FLOPs of different augmentation strategies. Unlike ChannelGAN which requires millions of parameters and billions of FLOPs, our proposed ADS+PRS only needs several thousand parameters and FLOPs to achieve a higher CSI recovery accuracy.

\begin{table}[!hbtp]
\renewcommand{\arraystretch}{1.5}
\caption{Parameters and computational complexity of augmentation strategies. B: Billion, M: Million, K: thousand.}
\begin{center}
\begin{tabular}{|c|c|c|c|c|}
\hline
\textbf{} & \textbf{ChannelGAN}  & \textbf{ADS} & \textbf{PRS} & \textbf{ADS+PRS} \\ \hline
\textbf{Parameters}     &  11.7 M   & 0.2 K & 32 & 0.2 K  \\ \hline
\textbf{FLOPs}      &   5.4 B  &  -  & 4.1 K & 4.1 K  \\ \hline
\end{tabular}\vspace*{-5mm}
\end{center}
\label{tab:aug-comp-complex}
\end{table}
\section{Conclusions}
This work develops a novel solution for training and deployment enhancement of DL models in massive MIMO CSI feedback. 
We consider two major obstacles to DL-based feedback frameworks:  new unseen channel environments and small training datasets 
from field measurement.
For new channel environments, 
we present an efficient scenario-adaptive CSI feedback architecture ``CSI-TransNet''. 
CSI-TransNet exploits the plug-in CSI-to-CSI translation module to reuse the pretrained anchor CSI model with high recovery accuracy and enables a lightweight encoder update in new scenarios. 
We develop an efficient deep unfolding-based CSI feedback network SPTM2-ISTANet+ as the CSI-TransNet backbone.
Against small measurement datasets,
we propose a simple and effective data augmentation strategy based on domain knowledge to replace blockbox GAN-based augmentation.
Our proposed augmentation strategy together with CSI-TransNet can significantly enhance CSI recovery performance with only a thousand encoder parameters 
for update and can achieve NMSE below $-20$ dB by using only $200$ measurement channel samples in three new scenarios at CR of ${1}/{4}$.

\textcolor{black}{Leveraging prior knowledge of channel variation proves instrumental in enhancing both data augmentation and the CSI feedback network. As we currently emphasize exploiting insights from wireless channel behaviors to refine CSI data augmentation and feedback network optimization without relying on auxiliary data, there remains an untapped potential. The dynamic nature of channels, characterized by their temporal and spatial correlations owing to shared local scattering clusters, offers prospects to reduce redundant information overhead in the time-varying channel. In the future, we aim to delve into precise channel variation modeling and the integration of side-information within model-driven DL networks, which can further optimize CSI feedback efficiency across varied environments.}

\section*{Acknowledgment}
The authors would like to acknowledge Yu-Chien Lin 
for his useful discussions in the process of preparing this manuscript.

\ifCLASSOPTIONcaptionsoff
  \newpage
\fi

\bibliographystyle{IEEEtran}
\bibliography{ref}

\end{document}